\documentclass[aps,prd,preprint,letterpaper,11pt,
			   amsmath, amssymb, amsfonts,floatfix,nofootinbib,
			   superscriptaddress]{revtex4-1}
\usepackage[utf8]{inputenc}
\usepackage[english]{babel}
\usepackage{microtype}
\usepackage{graphicx}
\usepackage{multirow}
\usepackage{dcolumn}
\usepackage{booktabs}
\usepackage{slashed}
\usepackage{lineno}
\usepackage{color}
\usepackage{hyperref}

\begin{document}

\preprint{FERMILAB-PUB-19-062-LBNF-ND}
\title{Neutrino tridents at DUNE}

\def\Cincy{Department of Physics, University of Cincinnati, Cincinnati, OH 45221, USA}
\def\harvard{Department of Physics, Harvard University, Cambridge, MA 02138, USA}
\def\ucsc{Santa Cruz Institute for Particle Physics, University of California, Santa Cruz, CA 95064, USA}

\author{Wolfgang~Altmannshofer} \email[]{waltmann@ucsc.edu}
\affiliation{\ucsc}

\author{Stefania~Gori} \email[]{sgori@ucsc.edu}
\affiliation{\ucsc}

\author{Justo~Mart\'in-Albo} \email[]{jmartinalbo@fas.harvard.edu}
\affiliation{\harvard}


\author{Alexandre Sousa} \email[]{alex.sousa@uc.edu} 
\affiliation{\Cincy}

\author{Michael Wallbank} \email[]{michael.wallbank@uc.edu} 
\affiliation{\Cincy}

\begin{abstract}
The DUNE near detector will collect an unprecedented large number of neutrino interactions, allowing the precise measurement of rare processes such as neutrino trident production, i.e.\ the generation of a lepton-antilepton pair through the scattering of a neutrino off a heavy nucleus. The event rate of this process is a powerful probe to a well-motivated parameter space of new physics beyond the Standard Model. In this paper, we perform a detailed study of the sensitivity of the DUNE near detector to neutrino tridents. We provide 
predictions for the Standard Model cross sections and corresponding event rates at the near detector for the $\nu_\mu \to \nu_\mu \mu^+\mu^-$, $\nu_\mu \to \nu_\mu e^+e^-$ and $\nu_\mu \to \nu_e e^+ \mu^-$ trident interactions (and the corresponding anti-neutrino modes), discussing their uncertainties. We analyze all relevant backgrounds, utilize a Geant4-based simulation of the DUNE-near detector liquid argon TPC (the official DUNE simulation at the time of writing this paper), and identify a set of selection cuts that would allow the DUNE near detector to measure the $\nu_\mu \to \nu_\mu \mu^+\mu^-$ cross section with a $\sim 40\%$ accuracy after running in neutrino and anti-neutrino modes for $\sim$3 years each. We show that this measurement would be highly sensitive to new physics, and, in particular, we find that the parameter space of models with gauged $L_\mu - L_\tau$ that can explain the $(g-2)_\mu$ anomaly could be covered with large significance. As a byproduct, a new Monte Carlo tool to generate neutrino trident events is made publicly available.
\end{abstract}

\maketitle
\tableofcontents

\section{Introduction}
Neutrino trident production is a weak process by which a neutrino, scattering off the Coulomb field of a heavy nucleus, generates a pair of charged leptons \cite{Czyz:1964zz,Lovseth:1971vv,Fujikawa:1971nx,Koike:1971tu,Koike:1971vg,Brown:1973ih,Belusevic:1987cw}. Measurements of muonic neutrino tridents, $\nu_\mu \to \nu_\mu \mu^+\mu^-$, were performed at the CHARM-II~\cite{Geiregat:1990gz}, CCFR~\cite{Mishra:1991bv} and NuTeV~\cite{Adams:1999mn} experiments:
\[ \label{eq:trident_exp}
\frac{\sigma(\nu_\mu \to \nu_\mu \mu^+\mu^-)_\text{exp}}{\sigma(\nu_\mu \to \nu_\mu \mu^+\mu^-)_\text{SM}} = 
\begin{cases}
1.58 \pm 0.64         & \text{(CHARM-II)} \\ 
0.82 \pm 0.28         & \text{(CCFR)} \\
0.72 ^{+1.73}_{-0.72} & \text{(NuTeV)} 
\end{cases}
\]
Both CHARM-II and CCFR found rates compatible with Standard Model (SM) expectations. No signal could be established at NuTeV. Future neutrino facilities, such as LBNF/DUNE \cite{Abi:2018dnh, Acciarri:2016crz, Acciarri:2016ooe, Acciarri:2015uup}, will offer excellent prospects to improve these measurements~\cite{Altmannshofer:2014pba,Magill:2016hgc,Ge:2017poy,Ballett:2018uuc}. A deviation from the event rate predicted by the SM could be an indication of new interactions mediated by new gauge bosons~\cite{Altmannshofer:2014pba}. This could happen, for example, if neutrinos were charged under new gauge symmetries beyond the SM gauge group, $SU(3)_c\times SU(2)_L\times U(1)_Y$.

In this paper, we study in detail the prospects for measuring neutrino trident production at the near detector of DUNE. As will be discussed below, the trident cross section is to a good approximation proportional to the charge squared ($Z^2$) of the target nuclei: $Z = 18$ for argon (DUNE), $Z = 14$ for silicon (CHARM II), and $Z = 26$ for iron (CCFR and NuTeV). As we will demonstrate, despite the smaller $Z^2$ compared to CCFR and NuTeV, the high-intensity muon-neutrino beam at the DUNE near detector leads to a sizable production rate of neutrino tridents.
The main challenge to obtain a precise measurement of the trident cross section is to distinguish the trident events from the copious backgrounds, mainly consisting of charged-current single-pion production events, $\nu_\mu N \to \mu \pi N^\prime$, as muon and pion tracks can be easily confused. Here, we identify a set of kinematic selection cuts that strongly suppress the background, allowing a measurement of the $\nu_\mu \to \nu_\mu \mu^+\mu^-$ cross section at DUNE.

Our paper is organized as follows. In Section~\ref{sec:SM}, we compute the cross sections for several neutrino-induced trident processes in the SM, and discuss the theoretical uncertainties in the calculation. We also provide the predicted event rates at the DUNE near detector. Section~\ref{sec:discovery} describes the sensitivity study. We analyze the kinematic distributions of signal and backgrounds, and determine the accuracy with which the $\nu_\mu \to \nu_\mu \mu^+\mu^-$ cross section can be measured at the DUNE near detector. In Section~\ref{sec:NP}, we analyze the impact that such a measurement will have on physics beyond the SM, both model independently and in the context of a $Z^\prime$ model with gauged $L_\mu - L_\tau$. We conclude in Section~\ref{sec:conclusions}. Details about nuclear and nucleon form factors and our implementation of the Borexino bound on the $Z^\prime$ parameter space are given in the Appendices~\ref{app:ff}, \ref{app:ff2}, and~\ref{app:borexino}. Our neutrino trident Monte Carlo generator tool can be found as an ancillary file on the \texttt{arXiv} entry for this paper.

\section{Neutrino tridents in the Standard Model} \label{sec:SM}

\subsection{SM predictions for the neutrino trident cross section} \label{sec:SM_Xsection}
\begin{figure}[tb!]
\centering
\includegraphics[width=0.35\textwidth]{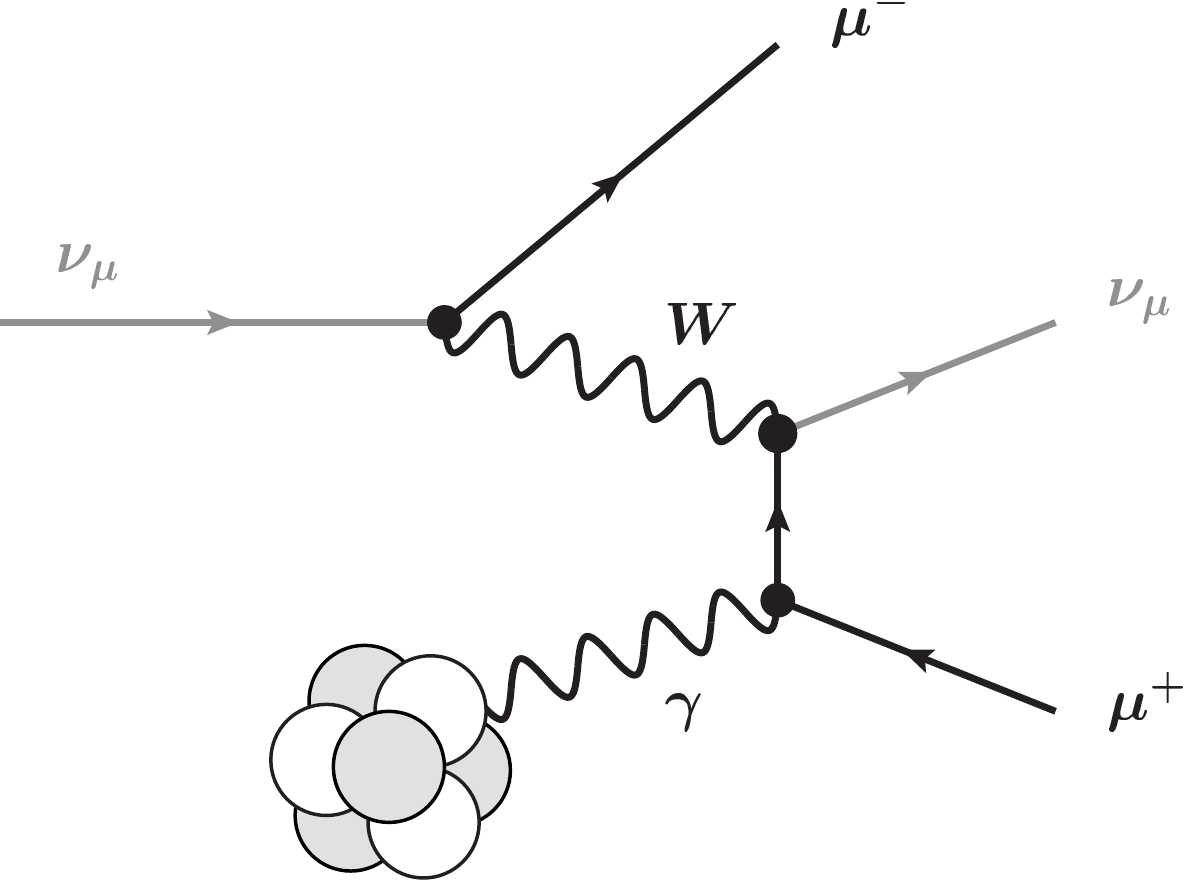} \qquad
\includegraphics[width=0.35\textwidth]{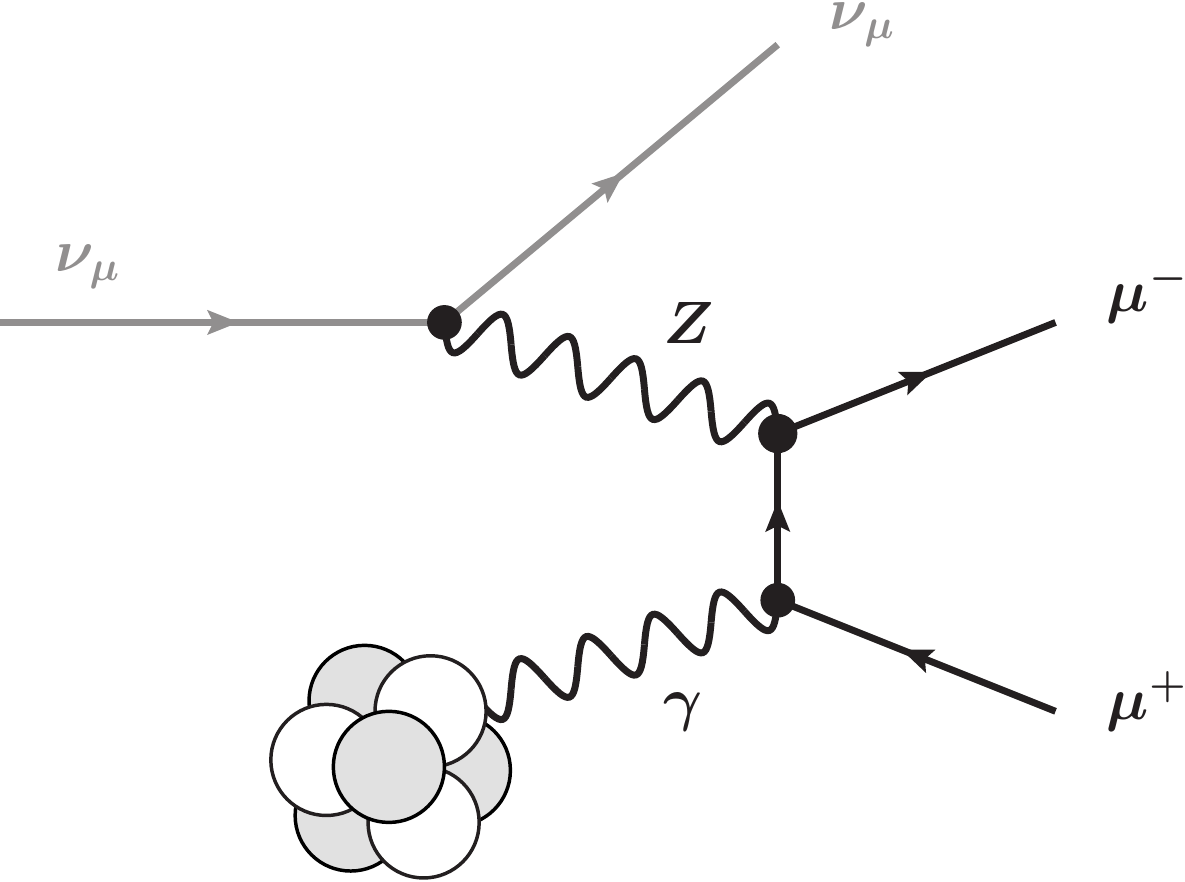} \\[\baselineskip]
\includegraphics[width=0.35\textwidth]{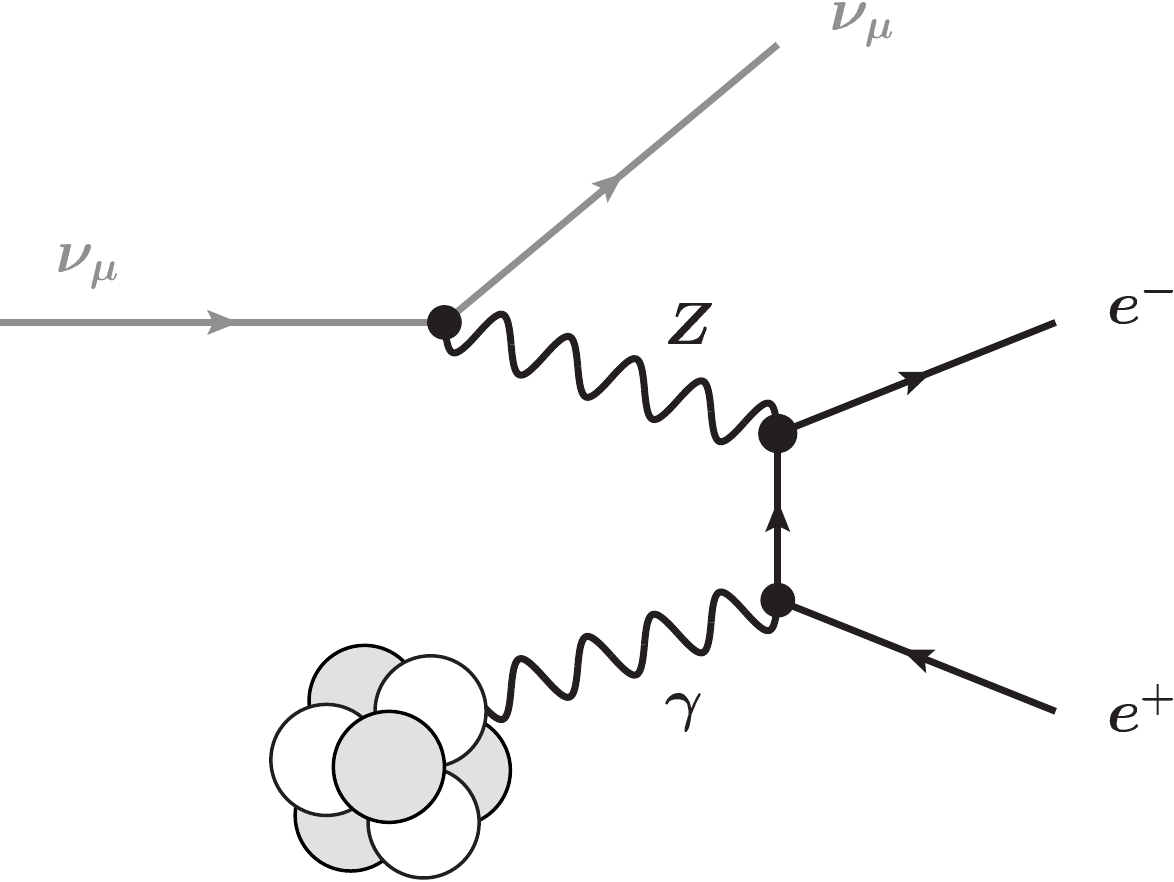} \qquad
\includegraphics[width=0.35\textwidth]{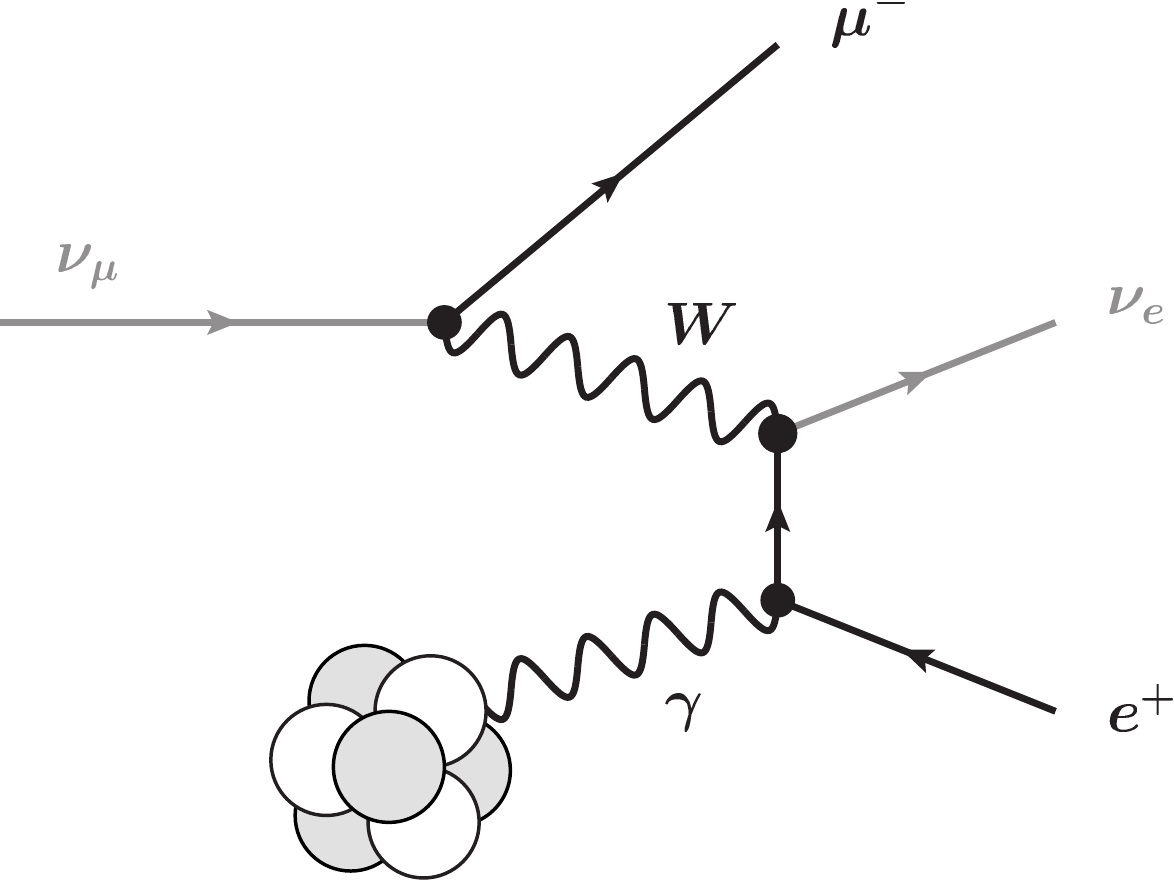} \\[\baselineskip]
\caption{Example diagrams for muon-neutrino-induced trident processes in the Standard Model. A second set of diagrams where the photon couples to the negatively charged leptons is not shown. 
Analogous diagrams exist for processes induced by different neutrino flavors and by anti-neutrinos.}
\label{fig:diagrams}
\end{figure}

Lepton-pair production through the scattering of a neutrino in the Coulomb field of a nucleus can proceed in the SM via the electro-weak interactions. Figure~\ref{fig:diagrams} shows example diagrams for various charged lepton flavor combinations that can be produced from a muon-neutrino in the initial state: a $\mu^+ \mu^-$ pair can be generated by $W$ and $Z$ exchange (top, left and right diagrams); an $e^+ e^-$ pair can be generated by $Z$ exchange (bottom left); an $e^+ \mu^-$ pair can be generated by $W$ exchange (bottom right). A muon neutrino cannot generate $\mu^+ e^-$ in the SM. Analogous processes can be induced by the other neutrino flavors and also by anti-neutrinos. The amplitude corresponding to the diagrams shown in the figure has a first-order dependence on the Fermi constant. Additional SM diagrams where the lepton system interacts with the nucleus through $W$ or $Z$ boson exchange instead of photon exchange are suppressed by higher powers of the Fermi constant and are therefore negligible.

\begin{figure}[tb!] 
\centering
\includegraphics[width=0.32\textwidth]{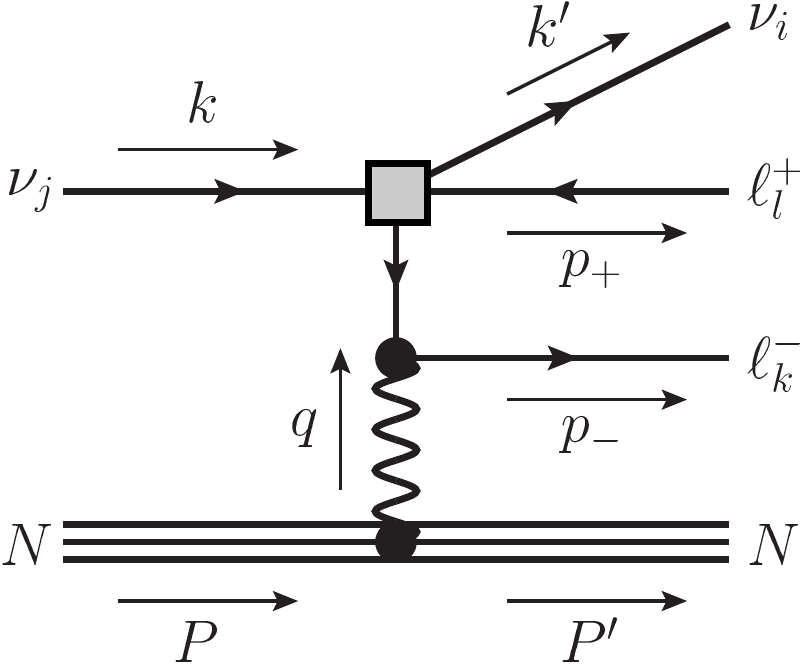} \qquad \qquad
\includegraphics[width=0.32\textwidth]{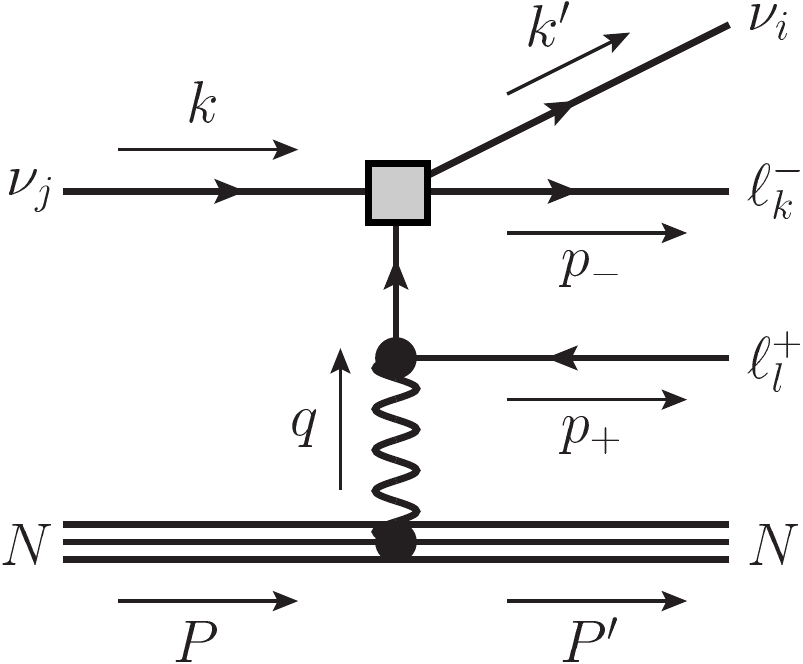}
\caption{Diagrams for the $\nu_j \to \nu_i \ell_k^- \ell_l^+$ trident process using the effective interaction of Eq.~(\ref{eq:Heff}).}
\label{fig:momenta}
\end{figure}

The weak gauge bosons of the SM are much heavier than the relevant momentum transfer in the trident process. Therefore, the effect of the $W$ and $Z$ bosons is accurately described by a four lepton contact interaction. After performing a Fierz transformation, the effective interaction can be written as 
\begin{equation} \label{eq:Heff}
 \mathcal H_\text{eff}^\text{SM} = \frac{G_F}{\sqrt{2}} \sum_{i,j,k,l} \left( g^V_{ijkl}~ (\bar \nu_i \gamma_\alpha P_L \nu_j)(\bar \ell_k \gamma^\alpha \ell_l) + g^A_{ijkl}~ (\bar \nu_i \gamma_\alpha P_L \nu_j)(\bar \ell_k \gamma^\alpha \gamma_5 \ell_l) \right) ~,
\end{equation}
with vector couplings $g^V$ and axial-vector couplings $g^A$. The indexes $i,j,k,l~(=e,\mu,\tau)$ denote the SM lepton flavors.
The values for the coefficients $g^V$ and $g^A$ for a variety of trident processes in the SM are listed in Table~\ref{tab:couplings}. These factors are the same as obtained in Ref.~\cite{Magill:2016hgc}.
Using the effective interactions, there are two Feynman diagrams that contribute to the trident processes. They are shown in Figure~\ref{fig:momenta}.

\begin{table}[tb!] 
\caption{Effective Standard Model vector and axial-vector couplings, as defined in Eq.~(\ref{eq:Heff}), for a variety of neutrino trident processes.}
\label{tab:couplings}
\begin{ruledtabular}
\begin{tabular}{lcc}
Process & $g^V_\text{SM}$ & $g^A_\text{SM}$ \\
\hline
$\nu_e \to \nu_e e^+e^-$ & $1 + 4 \sin^2\theta_W$ & $-1$ \\
$\nu_e \to \nu_e \mu^+\mu^-$ & $-1 + 4 \sin^2\theta_W$ & $+1$ \\
$\nu_e \to \nu_\mu \mu^+ e^-$ & $2$	& $-2$ \\
\hline
$\nu_\mu \to \nu_\mu e^+e^-$ & $-1 + 4 \sin^2\theta_W$ & $+1$ \\
$\nu_\mu \to \nu_\mu \mu^+\mu^-$ & $1 + 4 \sin^2\theta_W$ & $-1$ \\
$\nu_\mu \to \nu_e e^+\mu^-$ & $2$	& $-2$ \\
\end{tabular}
\end{ruledtabular}
\end{table}

Given the above effective interactions, the cross sections for the trident processes can be computed in a straightforward way. The dominant contributions arise from the coherent elastic scattering of the leptonic system on the full nucleus. We will also consider incoherent contributions from elastic scattering on individual nucleons (referred to as \emph{diffractive scattering} in Refs.~\cite{Magill:2016hgc,Ballett:2018uuc}). 

In addition to the elastic scattering on the full nucleus or on individual nucleons, also inelastic processes can contribute to trident production. Inelastic processes include events where the nucleus scatters into an excited state, the excitation of a nucleon resonance, and deep-inelastic scattering. As shown in~\cite{Magill:2016hgc}, deep-inelastic scattering is negligible for trident production at the SHiP experiment. We expect inelastic processes to be negligible for the neutrino energies we consider.

All our results shown below are based on a calculation of the full $2 \to 4$ scattering process. We do not use the equivalent photon approximation that has been shown not to be reliable in some cases~\cite{Ballett:2018uuc}.

\subsubsection{Coherent scattering on nuclei}
The differential cross section of the coherent scattering process on a nucleus of mass $m_N$ is enhanced by $Z^2$ and can be expressed as~\cite{Lovseth:1971vv,Brown:1973ih} (see also \cite{Ge:2017poy,Ballett:2018uuc})
\begin{equation} \label{eq:dsigma_N}
\mathrm{d}\sigma_\text{coh.} = \frac{Z^2 \alpha_\text{em}^2 G_F^2}{128 \pi^6} \frac{1}{m_N E_\nu} \frac{\mathrm{d}^3k'}{2 E_{k'}}  \frac{\mathrm{d}^3p_+}{2 E_+}  \frac{\mathrm{d}^3p_-}{2 E_-}  \frac{\mathrm{d}^3P'}{2 E_{P'}} \frac{H^{\alpha\beta}_N L_{\alpha \beta}}{q^4} \delta^{(4)}(k - k' - p_+ - p_- + q) ~,
\end{equation}
where the momenta of the incoming and outgoing particles are defined in Fig.~\ref{fig:momenta} and $E_\nu$ is the energy of the incoming neutrino. 
The leptonic tensor $L_{\alpha \beta}$ is given by
\begin{eqnarray}
 L_{\alpha\beta} &=& \sum_{s,s',s_+,s_-} A_\alpha A_\beta^\dagger  ~, \nonumber \\ 
\text{with}~~ A_\alpha &=& (\bar u' \gamma_\mu P_L u) \left(\bar u_- \left[ \gamma_\alpha \frac{ p\!\!\!/_- - q\!\!\!/ + m_- }{(p_--q)^2 - m_-^2} \gamma^\mu (g^V_{ijkl} + g^A_{ijkl} \gamma_5) \right.\right. \nonumber \\
&& ~~~~~~~~~~~~~~~~~~~ \left.\left.- \gamma^\mu (g^V_{ijkl} + g^A_{ijkl} \gamma_5) \frac{ p\!\!\!/_+ - q\!\!\!/ + m_+ }{(p_+-q)^2 - m_+^2} \gamma_\alpha\right] v_+ \right)~,  \label{eq:L_ab}
\end{eqnarray}
where $m_\pm$, $s_\pm$, and $v_+$, $u_-$ are the masses, spins and spinors of the positively and negatively charged leptons and $s$, $s'$ and $u$, $u'$ are the spins and spinors of the incoming and outgoing neutrinos.

The relevant part of the hadronic tensor for coherent scattering on a spin $0$ nucleus is
\begin{equation}
 H^{\alpha\beta}_N = 4 P_\alpha P_\beta \big[F_N(q^2)\big]^2 ~,
\end{equation}
where $F_N(q^2)$ is the electric form factor of the nucleus, $N$, and $P$ the initial momentum of the nucleus. We use nuclear form factors based on measured charge distributions of nuclei~\cite{DeJager:1987qc}. Details about the nuclear form factors are given in the Appendix~\ref{app:ff}. 

The experimental signature of the coherent scattering are two opposite sign leptons without any additional hadronic activity.

\subsubsection{Incoherent scattering on individual nucleons}
\label{sec:incoherent}
In addition to the coherent scattering on the nucleus, the leptonic system can also scatter on individual nucleons inside the nucleus. The corresponding differential cross sections have a similar form and read
\begin{equation} \label{eq:dsigma_pn}
 \mathrm{d}\sigma_{p(n)} = \frac{\alpha_\text{em}^2 G_F^2}{128 \pi^6}\frac{1}{m_{p(n)} E_\nu} \frac{\mathrm{d}^3k'}{2 E_{k'}}  \frac{\mathrm{d}^3p_+}{2 E_+}  \frac{\mathrm{d}^3p_-}{2 E_-}  \frac{\mathrm{d}^3P'}{2 E_{P'}} \frac{H^{\alpha\beta}_{p(n)} L_{\alpha \beta}}{q^4} \delta^{(4)}(k - k' - p_+ - p_- + q) ~. 
\end{equation}
The leptonic tensor is still given by Eq.~(\ref{eq:L_ab}). The relevant part of the hadronic tensor for scattering on the spin 1/2 protons (neutrons) is
\begin{equation}
 H^{\alpha\beta}_{p(n)} = 4 P^\alpha P^\beta \left(\frac{4m_{p(n)}^2 \big[G_E^{p(n)}(q^2)\big]^2}{q^2 + 4m_{p(n)}^2} + \frac{q^2 \big[G_M^{p(n)}(q^2)\big]^2}{q^2 + 4m_{p(n)}^2} \right) + g^{\alpha\beta} q^2 \big[G_M^{p(n)}(q^2)\big]^2 ~,
\end{equation}
where $G_E^{p(n)}(q^2)$ and $G_M^{p(n)}(q^2)$ are the electric and magnetic form factor of the proton (neutron) and $m_{p(n)}$ is the proton (neutron) mass.
In our numerical calculations, we use form factors from a fit to electron-proton and electron-nucleus scattering data~\cite{Alberico:2008sz}. Details about the nucleon form factors are given in the Appendix~\ref{app:ff2}. 

The differential trident cross section corresponding to the incoherent processes is
\begin{equation}
\mathrm{d}\sigma_\text{incoh.} = \Theta(|\vec q|) \big( Z~\mathrm{d}\sigma_p + (A-Z)~\mathrm{d}\sigma_n \big) ~,
\end{equation}
where $Z$ and $(A-Z)$ are the number of protons and neutrons inside the nucleus, respectively. We include a Pauli blocking factor derived from the ideal Fermi gas model of the nucleus~\cite{Lovseth:1971vv}
\begin{equation}
\Theta(|\vec q|) = 
\begin{cases} 
\frac{3|\vec q|}{4 p_F} - \frac{|\vec q|^3}{16 p_F^3} \,, & \text{for}~|\vec q| < 2p_F \\ 
1\,, & \text{for}~|\vec q| > 2p_F
\end{cases},
\end{equation}
with the Fermi momentum $p_F = 235$~MeV and $\vec q$ the spatial component of the momentum transfer to the nucleus.

In addition to the two opposite sign leptons, the final state now contains an additional proton (or neutron) that is kicked out from the nucleus during the scattering process.

\subsubsection{Results for the cross section and discussion of uncertainties}\label{sec:resultsAndUncertainties}
To obtain the total cross sections for the coherent and incoherent processes discussed above, we integrate the four-particle phase space in~(\ref{eq:dsigma_N}) and~(\ref{eq:dsigma_pn}) numerically. Using the optimized integration variables identified in~\cite{Lovseth:1971vv}, we find that the numerical integration converges reasonably fast. We checked explicitly that our numerical computation accurately reproduces the cross section tables for a set of fixed neutrino energies given in~\cite{Lovseth:1971vv}. We estimate the uncertainty of our numerical integration procedure to be around the per-mille level, which is negligible compared to the other uncertainties discussed below.

\begin{figure}[tb!]
\centering
\includegraphics[width=0.48\textwidth]{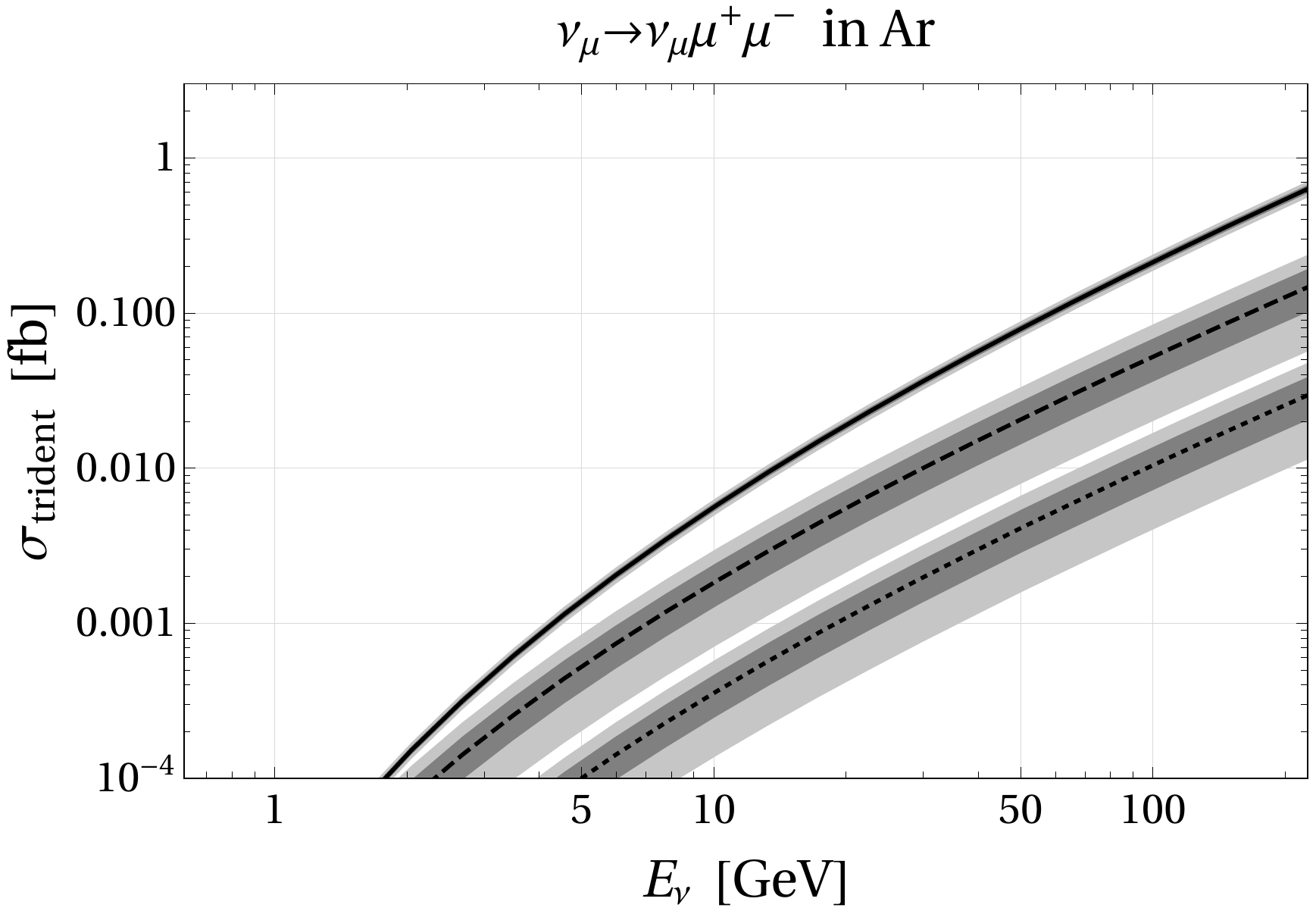}
\includegraphics[width=0.48\textwidth]{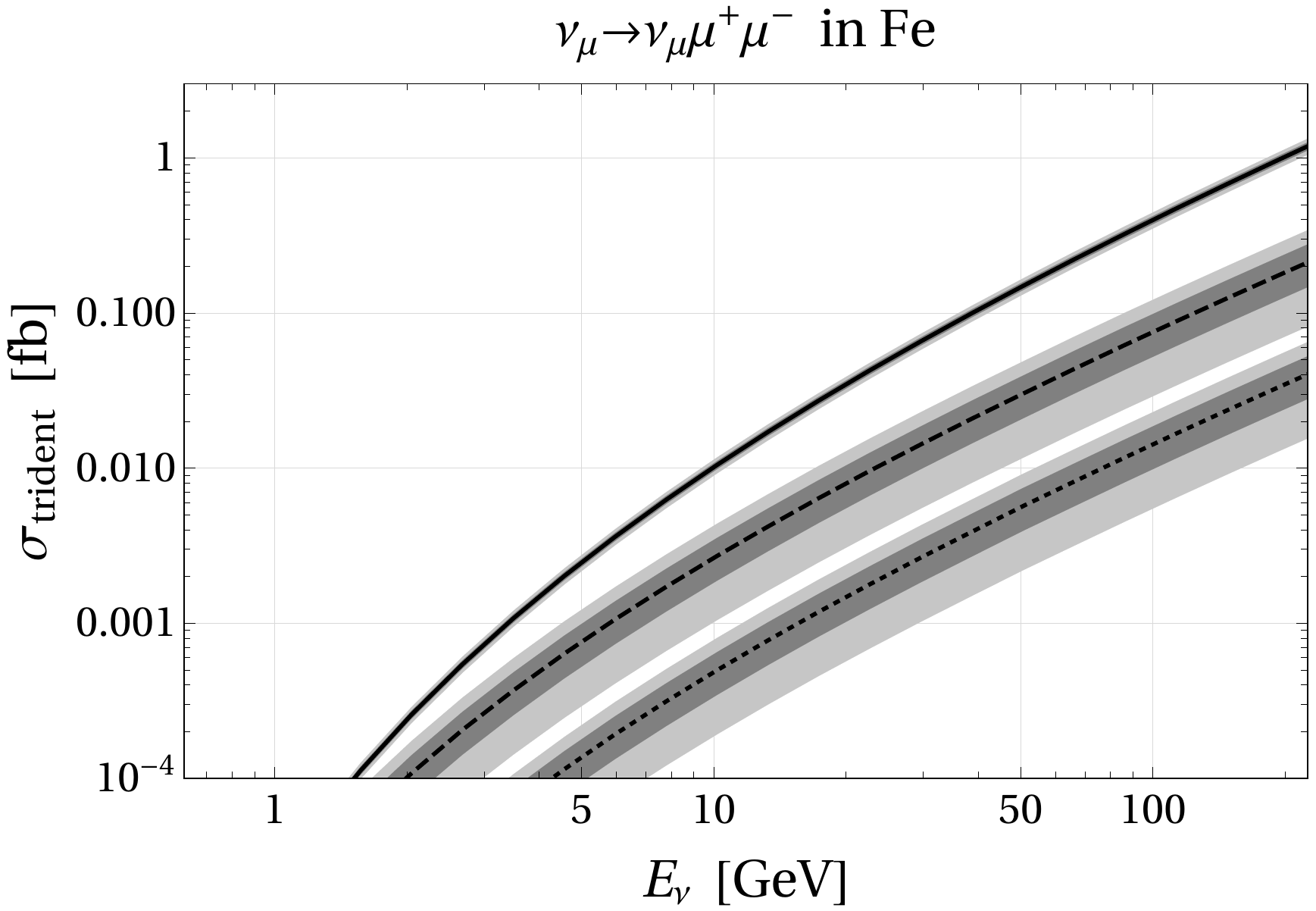} 
\\[0.5\baselineskip]
\includegraphics[width=0.48\textwidth]{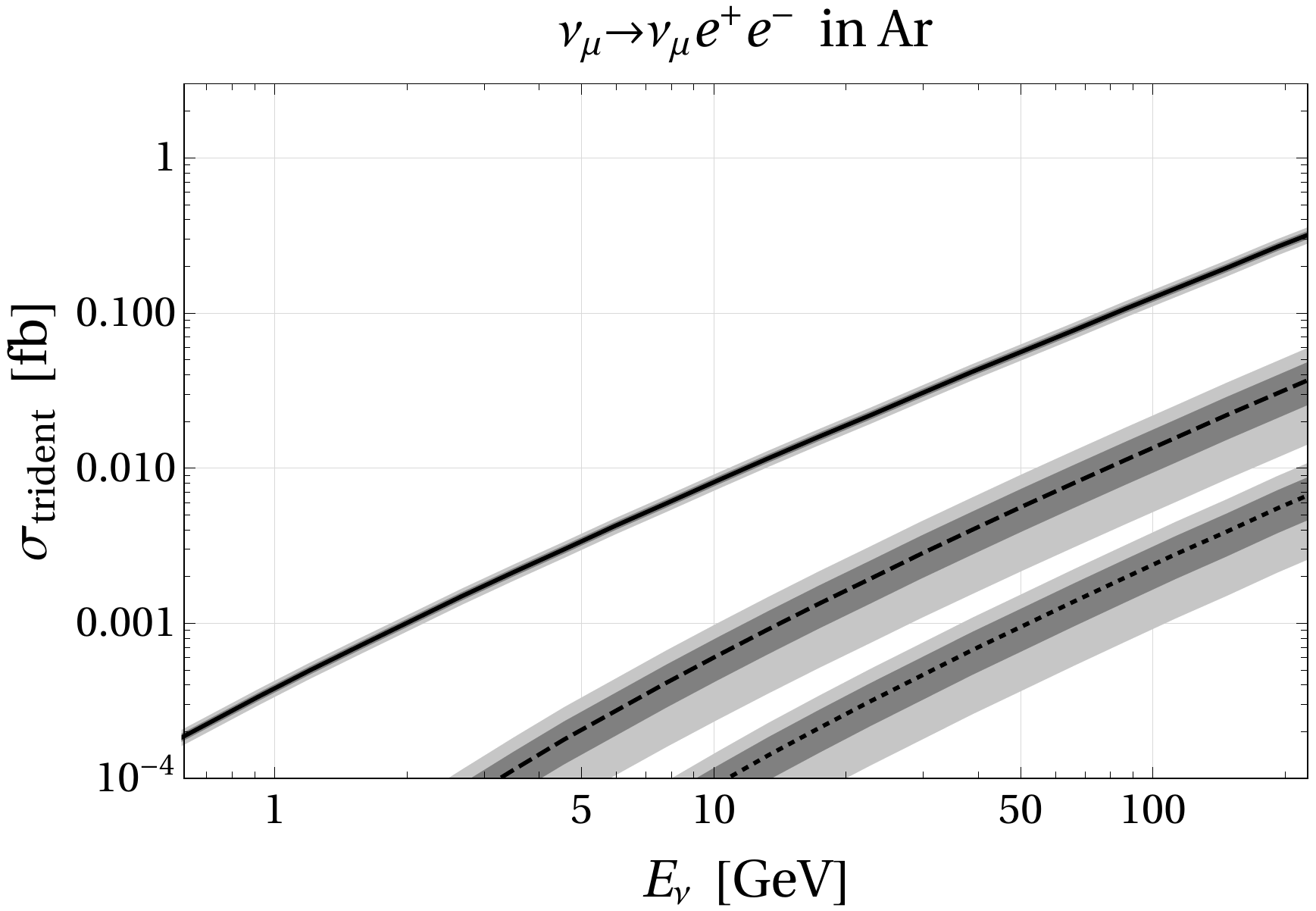}
\includegraphics[width=0.48\textwidth]{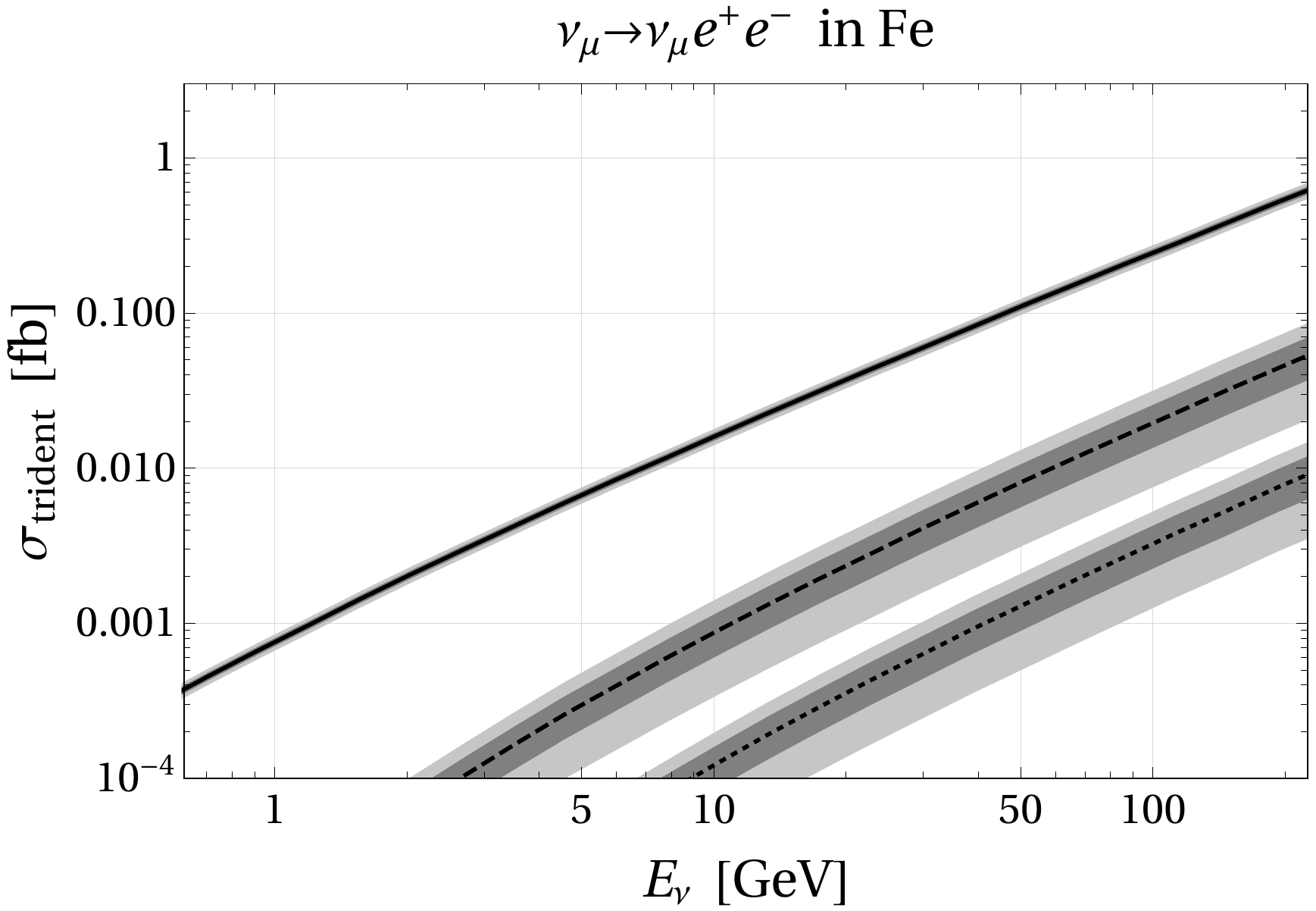}
\\[0.5\baselineskip]
\includegraphics[width=0.48\textwidth]{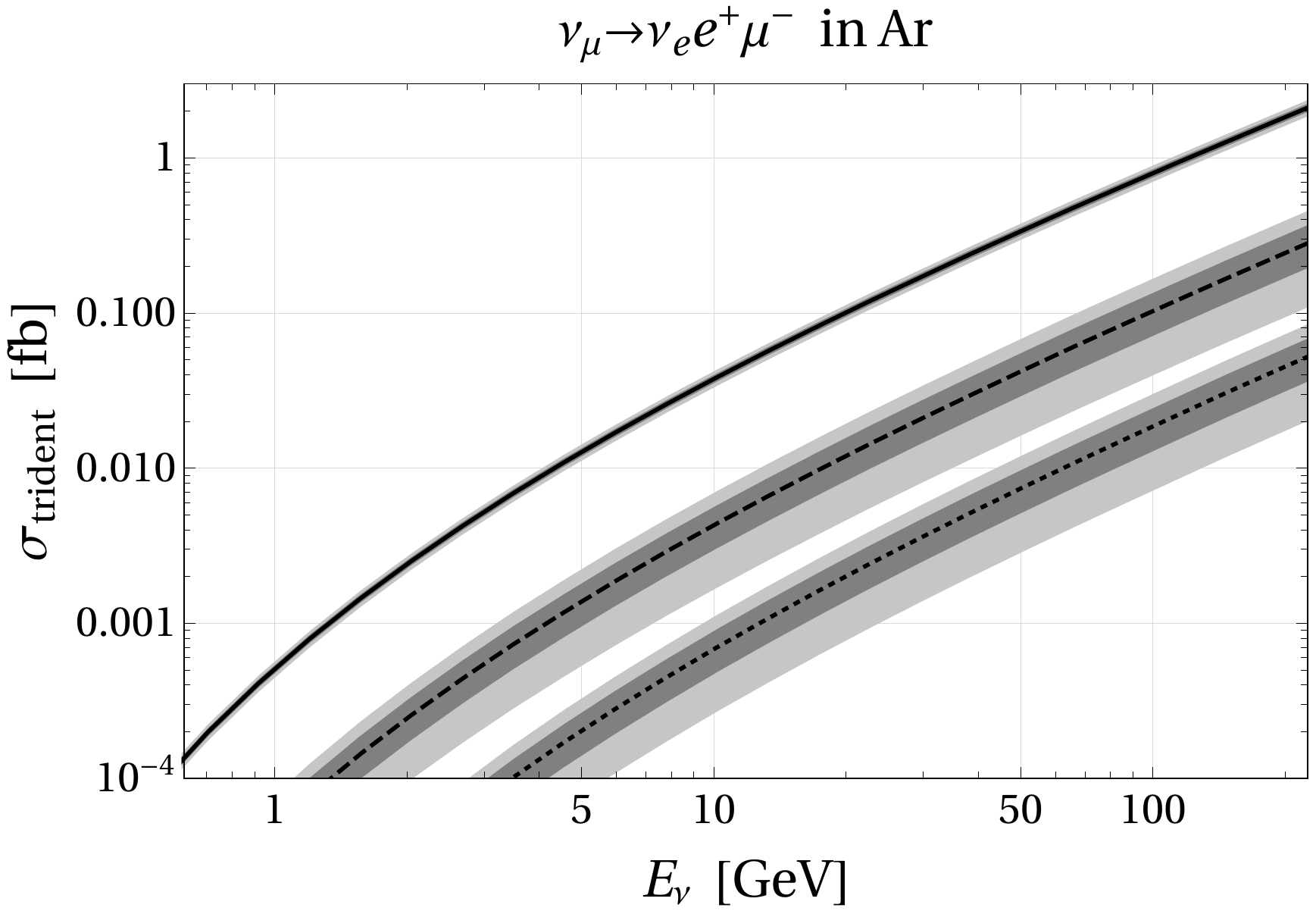}
\includegraphics[width=0.48\textwidth]{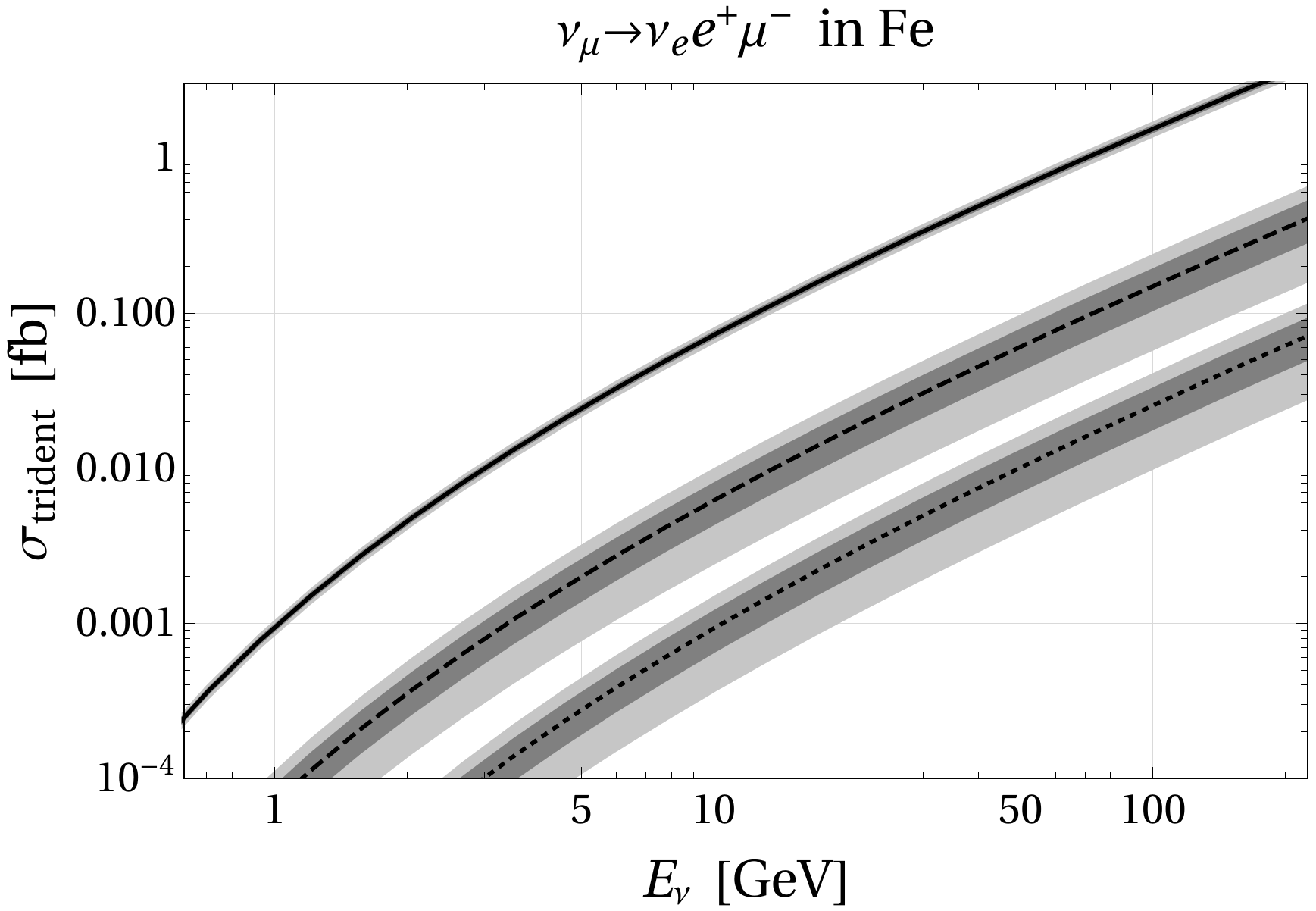}
\caption{Standard Model predictions for the cross sections of the trident processes $\nu_\mu \to \nu_\mu \mu^+\mu^-$, $\nu_\mu \to \nu_\mu e^+e^-$, and $\nu_\mu \to \nu_e e^+\mu^-$ for scattering on argon (left) and iron (right) as a function of the energy of the incoming neutrino. Shown are both the coherent component (solid) and incoherent components (dashed for proton, dotted for neutron). The $1\sigma$ and $2\sigma$ cross-section uncertainties are indicated by the shaded bands.}
\label{fig:Xsection}
\end{figure}

In Figure~\ref{fig:Xsection} we show the cross sections for the $\nu_\mu \to \nu_\mu \mu^+\mu^-$, $\nu_\mu \to \nu_\mu e^+e^-$, and $\nu_\mu \to \nu_e e^+\mu^-$ processes for scattering on argon (left) and iron (right) as a function of the energy of the incoming neutrino. We show both the coherent and incoherent components. 
From the figure, we make the following observations:
\begin{itemize}
 \item The cross sections fall steeply at low neutrino energy, as it becomes more and more difficult to produce the lepton pair via scattering with a low $q^2$ photon from the Coulomb field.  
 \item The $\nu_\mu\to\nu_e\mu^+e^-$ process has the largest cross section over a broad range of neutrino energies since it arises from a $W$ mediated diagram (see Fig. \ref{fig:diagrams}). The cross section for the $\nu_\mu\to\nu_e e^+e^-$ process is smaller due to the smaller couplings of the $Z$ boson with leptons and neutrinos. The $\nu_\mu\to\nu_\mu\mu^+\mu^-$ process typically leads to the smallest cross section at low energies, due to destructive interference between the $W$ and the $Z$ contributions and the relatively heavy di-muon pair in the final state.
 \item For processes involving electrons, the incoherent cross section is approximately $5\% - 10\%$ of the coherent cross section.
For the $\nu_\mu \to \nu_\mu \mu^+\mu^-$ process, however, about $30\%$ of the cross section is coming from incoherent scattering. Scattering on individual protons and neutrons provides photons with higher $q^2$, which makes it easier to produce the (relatively) heavy di-muon pair.
 \item Among the incoherent processes, the cross section for scattering on protons is approximately one order of magnitude larger than for scattering on neutrons because neutrons are electrically neutral.
\end{itemize}

In Fig.~\ref{fig:Xsection}, estimates of the $1\sigma$ and $2\sigma$ uncertainties of the cross sections are indicated by the shaded bands. We consider uncertainties from form factors, higher order QED corrections, higher order weak corrections, and nuclear modeling.

Form factor uncertainties for the coherent scattering appear to be well under control (see Appendix~\ref{app:ff}). In our numerical analysis, we use $1\%$ uncertainty on the coherent cross section coming from form factors. For the incoherent scattering we find differences in the cross section of a few percent, using different nucleon form factors (see Appendix~\ref{app:ff2}). For our numerical analysis, we assign a $3\%$ value to the uncertainty arising from form factors.

Higher order QED effects might lead to non-negligible corrections. Naively, we estimate that the effects could be of order $Z \alpha_\text{em} / 4\pi \simeq 1\%$ for argon. Moreover, at tree level there is ambiguity about which value of $\alpha_\text{em}$ should be used in the cross section computation. The $q^2$ of the photon from the Coulomb field is typically very low, suggesting that the zero momentum value should be appropriate. We use $\alpha_\text{em} = 1/137$ and conservatively assign a $3\%$ uncertainty to the total cross section from higher order QED effects.

At lowest order in the weak interactions, the value for the weak mixing angle is ambiguous. Using the on-shell value $\sin^2\theta_W = 0.22336$~\cite{Patrignani:2016xqp} or the $\overline{\text{MS}}$ value at zero momentum transfer $\sin^2\theta_W = 0.23868$~\cite{Erler:2017knj}, it leads, for example, to a $\sim 5\%$ shift in the cross section of the $\nu_\mu \to \nu_\mu \mu^+\mu^-$ process. In our numerical analysis we use the $\overline{\text{MS}}$ value at the electro-weak scale $\sin^2\theta_W = 0.23129$~\cite{Patrignani:2016xqp} and assign a $5\%$ uncertainty due to higher order weak corrections to all cross sections.

A large uncertainty might originate from the nuclear modeling of the incoherent processes.
As described above, we include a Pauli blocking factor that is derived by treating the nucleus as an ideal Fermi gas. In~\cite{LlewellynSmith:1971uhs}, differences in incoherent scattering cross sections of $\mathcal O(20\%)$ are found by comparing the Fermi gas model with more sophisticated shell models. As additional effects like rescattering or absorption of the nucleon in the nucleus might further modify the cross section, we use 30\% uncertainty on all incoherent cross sections to be conservative.
A more sophisticated nuclear model would be required to obtain a more precise prediction of the incoherent cross sections. 

To determine the total uncertainty we add all individual uncertainties in quadrature.
The final uncertainties on the coherent cross sections that we find are approximately $6\%$ and they are dominated by our estimate of possible higher order electro-weak corrections. For the incoherent scattering cross sections, the by far dominant uncertainty is due to the nuclear modeling.

\subsection{Neutrino tridents at the DUNE near detector}\label{sec.DUNEsigma}

\subsubsection{The DUNE near detector}
The \emph{Deep Underground Neutrino Experiment} (DUNE) \cite{Abi:2018dnh, Acciarri:2016crz, Acciarri:2016ooe, Acciarri:2015uup} is an international project for neutrino physics and nucleon-decay searches, currently in the design and planning stages. Once built, DUNE will consist of two detectors exposed to a megawatt-scale, wide-band muon-neutrino beam produced at the Fermi National Accelerator Laboratory (Illinois, USA). One of the detectors will record neutrino interactions near the beginning of the beamline, while the other, much larger, detector, comprising four 10-kilotonne liquid argon time projection chambers (TPCs), will be installed at a depth of 1.5 km at the Sanford Underground Research Facility (South Dakota, USA), about 1300~kilometres away of the neutrino source. Among the primary scientific goals of DUNE are the precision measurement of the parameters that govern neutrino mixing ---\thinspace including those still unknown: the octant in which the $\theta_{23}$ mixing angle lies, the neutrino mass ordering and the value of the CP-violation phase\thinspace---, as well as nucleon-decay searches and neutrino astrophysics.

One of the main roles of the DUNE near detector (ND), which will be located 570~meters away from the beamline production target, is the precise characterization of the neutrino beam energy and composition, as well as the measurement to unprecedented accuracy of the cross sections and particle yields of the various neutrino scattering processes. Additionally, as the ND will be exposed to an intense flux of neutrinos, it will collect an extraordinarily large sample of neutrino interactions, allowing for an extended science program that includes searches for new physics (e.g.\ heavy sterile neutrinos or non-standard interactions). 

The DUNE ND is presently under design. The baseline detector concept consists of a liquid argon TPC (LArTPC) and a magnetized high-resolution tracker \cite{DUNE-ND_CDR:2018}. The latter, not considered for the study discussed in this paper, will consist of a large high-pressure argon gas TPC surrounded by an electromagnetic sampling calorimeter. The design of the LArTPC will be based on the ArgonCube concept \cite{Amsler:1993255}, which places identical but separate TPC modules in a common bath of liquid argon. Each module features a central cathode and two drift volumes with pixelized charge readouts and light detection systems. Module walls are kept thin to provide transparency to the tracks and showers produced in neutrino interactions. This detector configuration will mitigate the effects of event pile-up and allow for an optimal use of liquid argon by boasting a relatively large active volume. The dimensions presently considered for the LArTPC, imposed by requirements on event statistics and containment, are 7~m width, 3~m height and 5~m depth, corresponding to an argon mass of about 147~tonnes. The definitive DUNE ND configuration will be defined in an upcoming near detector \emph{Conceptual Design Report} (CDR) and in a subsequent \emph{Technical Design Report} (TDR).

\subsubsection{Expected event rates in the Standard Model}
In Table~\ref{tab:events} we show the number of expected events of muon-neutrino-induced Standard Model trident events at the DUNE near detector per tonne of argon and year of operation in the neutrino-beam (first four rows) or antineutrino-beam (last four rows) configurations. Note that the number of events for the incoherent process is mainly coming from the scattering with protons. As discussed in Sec.~\ref{sec:incoherent}, the neutron contribution (included as well in the table) is much smaller and amounts to only $\sim 10\%$ of the total incoherent cross section. In parenthesis, we also show the number of expected events for 147 tonnes of argon and a run of 3~years.

\begin{table}[tb]
\caption{Expected number of muon-neutrino-induced Standard Model trident events at the DUNE near detector per tonne of argon and year of operation in neutrino mode (first four rows) or anti-neutrino mode (last four rows). The numbers in parenthesis correspond to the total statistics in the 147-tonne LArTPC for a run of 3~years.} \label{tab:events}
\begin{center}
\begin{tabular}{@{\hspace{2.5ex}}lcc}
\hline\hline
& ~~~Coherent~~~ & ~~~Incoherent~~~ \\ \hline
\multirow{2}{*}{$\nu_\mu \to \nu_\mu\ \mu^+\mu^-$} & $1.17 \pm 0.07$ & $0.49 \pm 0.15$ \\
 & $(516 \pm 31)$ & $(216 \pm 66)$ \\
\multirow{2}{*}{$\nu_\mu \to \nu_\mu\ e^+e^-$} & $2.84 \pm 0.17$ & $0.18 \pm 0.06$ \\
 & $(1252 \pm 75)$ & $(79 \pm 27)$ \\
\multirow{2}{*}{$\nu_\mu \to \nu_e\ e^+\mu^-$} & $9.8 \pm 0.6$ & $1.2 \pm 0.4$ \\
& $(4322 \pm 265)$ & $(529 \pm 176)$ \\
\multirow{2}{*}{$\nu_\mu \to \nu_e\ \mu^+e^-$} & $0$ & $0$ \\
& $(0)$ & $(0)$ \\ \hline
\multirow{2}{*}{$\bar\nu_\mu \to \bar\nu_\mu\ \mu^+\mu^-$} & $0.72 \pm 0.04$ & $0.32 \pm 0.10$ \\
& $(318 \pm 18)$ & $(141 \pm 44)$ \\
\multirow{2}{*}{$\bar\nu_\mu \to \bar\nu_\mu\ e^+e^-$} & $2.21 \pm 0.13$ & $0.13 \pm 0.04$ \\
& $(975 \pm 57)$ & $(57 \pm 18)$ \\
\multirow{2}{*}{$\bar\nu_\mu \to \bar\nu_e\ e^+\mu^-$} & $0$ & $0$ \\
& $(0)$ & $(0)$ \\
\multirow{2}{*}{$\bar\nu_\mu \to \bar\nu_e\ \mu^+e^-$} & $7.0 \pm 0.4$ & $0.9 \pm 0.3$ \\
& $(3087\pm176)$ & $(397\pm132)$ \\
\hline\hline
\end{tabular}
\end{center}
\end{table}

\begin{figure}[!tb]
\centering
\includegraphics[width=0.48\textwidth]{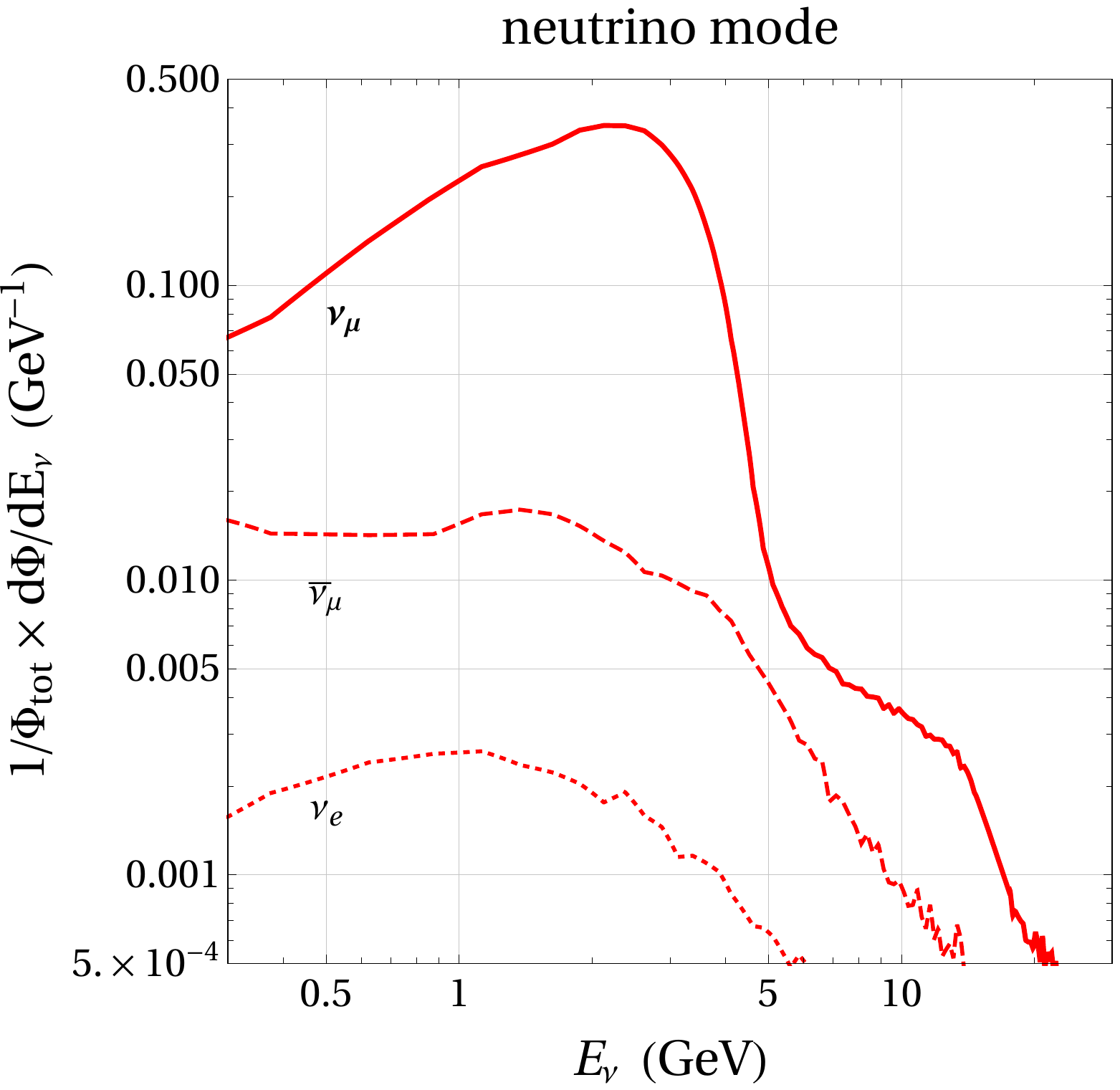} \quad \includegraphics[width=0.48\textwidth]{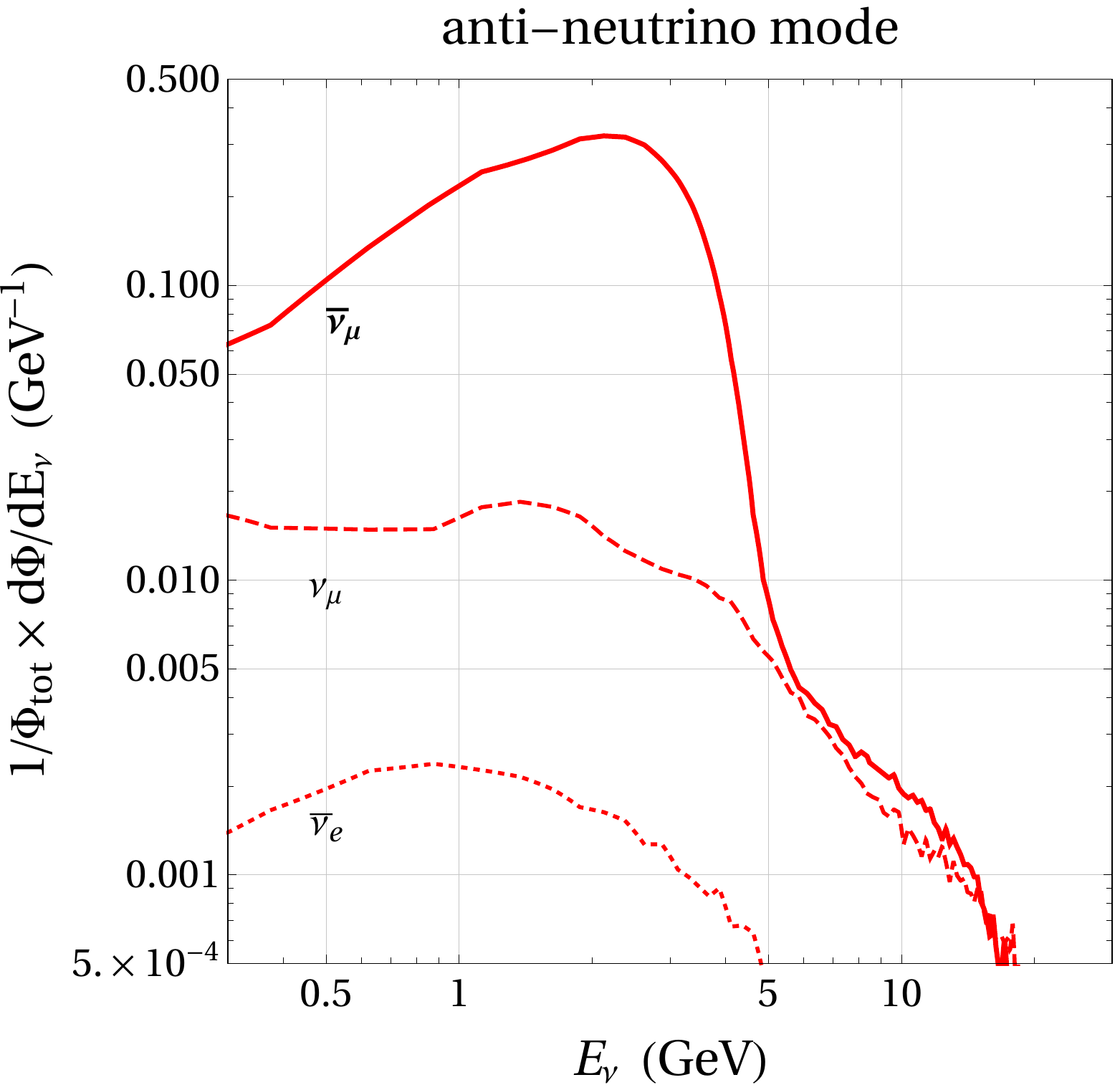}
\caption{Normalized energy spectra of the neutrino species at the DUNE near detector for the neutrino-beam (left) and antineutrino-beam (right) modes of operation.}
\label{fig:neutrinoEnergies}
\end{figure}

The normalized neutrino beam energy spectra are shown in Fig.~\ref{fig:neutrinoEnergies}. The relevant integrated flux in neutrino mode is $F_{\nu_\mu} = 1.04 \times 10^{-3}~ m^{-2}~ \text{POT}^{-1} $ and in antineutrino mode $\bar F_{\bar \nu_\mu} = 0.94 \times 10^{-3}~ m^{-2}~ \text{POT}^{-1} $~\cite{dune}.
We assume $1.1 \times 10^{21}$~POT per year. 

The numbers of events for the coherent process in Tab.~\ref{tab:events} are then obtained via
\begin{equation}
N_{\rm{trident}}= F~\sigma~N_{\rm{Ar}}~N_{\rm{POT}}=F~\sigma~\frac{M_D}{M_{\rm{Ar}}}~N_{\rm{POT}},
\end{equation}
where $F$ is the relevant integrated neutrino flux as given above, $\sigma$ is the neutrino trident cross section (convoluted with the corresponding normalized energy distribution), $N_{\rm{Ar}}$ the number of argon nuclei in the detector, $M_D$ the mass of the detector, $M_{\rm{Ar}}$ the mass of argon ($39.95u$), and $N_\text{POT}$ is the number of protons on target. Similarly, we computed the number of events for the incoherent processes.

In our calculation of the number of trident events we neglect all flux components but the $\nu_\mu$ component in neutrino mode and the $\bar \nu_\mu$ component in anti-neutrino mode. Taking into account the scattering of the other components will increase the expected event numbers by a few percent, which is within the given uncertainties. The rates for the antineutrino-beam mode are smaller by approximately 30\%, mainly due to the lower flux.



\section{Discovering SM muon tridents at DUNE} \label{sec:discovery}
In this section, we discuss the prospects for detecting muon trident events, $\nu_\mu \to \nu_\mu~\mu^+~\mu^-$, at the DUNE near detector. As we will discuss in Section~\ref{sec:NP}, this process is particularly relevant to test new light gauge bosons that couple to second generation leptons. A detailed discussion of electron tridents and electron-muon tridents at the DUNE near detector is left for future work. 

\subsection{Simulation}
The study presented here makes use of Monte-Carlo datasets generated with the official (at the time of writing of the paper) DUNE Geant4 \cite{Agostinelli:2002hh} simulation of the ND LArTPC. Each simulated event represents a different neutrino-argon interaction in the active volume of the detector. All final-state particles produced in the interactions are propagated by Geant4 through the detector geometry until they deposit all their energy or leave its boundaries. In this process, additional particles (which are tracked as well) may be generated via scattering or decay. The trajectories and associated energy deposits left by charged particles in the active volume of the LArTPC are recorded and written to an output file. 

For simplicity, charge collection and readout are not simulated, but their effect on the data is taken into account in our study with the introduction of the typical detection thresholds and resolutions expected from the ND LArTPC. Given that state-of-the-art TPCs have achieved very high reconstruction efficiency ($>90\%$) in significantly busier environments (e.g.\ the ALICE experiment \cite{ALICE:2013}), we neglect the effect of mis-reconstructed events. Likewise, we ignore the possible backgrounds or the inefficiency arising from interaction pile-up (i.e. the cross-contamination of different neutrino interactions occurring in the same TPC event) since the detector design will be optimized to make it negligible \cite{DUNE-ND_CDR:2018}.

Muon trident signal events are generated using the standalone Monte Carlo event generator that we have written and that simulates muon-neutrino and electron-neutrino induced trident events through the scattering off argon and iron nuclei. Neutrino fluxes of the CCFR experiment and the DUNE experiment are implemented. The phase space sampling is based on the optimized kinematical variables that were identified in Ref.~\cite{Lovseth:1971vv}. The \verb|C++| source code of the event generator is publicly available as an ancillary file on the \href{https://arxiv.org/abs/1902.06765}{arXiv}.

Several SM processes can constitute background for the muon trident process. In our simulation, we generate $10^8$ neutrino interactions using the GENIE Monte Carlo generator~\cite{Andreopoulos:2009rq,Andreopoulos:2015wxa}.
By far, the most important background is due to the mis-identification of charged-pion tracks. Roughly $38\%$ of the events have a charged lepton and a charged pion in the final state, leading to two muon-like charged tracks, as in our trident signal.  We find that di-muon events from charged current charm production only represent less than one percent of the total background.

\subsection{Kinematic distributions and event selection}
We identify a set of optimal kinematic variables that help discriminating between signal and background. Particularly, we use the number of tracks, the angle between tracks, the length of the tracks, and the total energy deposited within 10 cm of the neutrino interaction vertex ($E_{10}$). 

\begin{figure}[tb]
\centering
\includegraphics[width=0.441\textwidth]{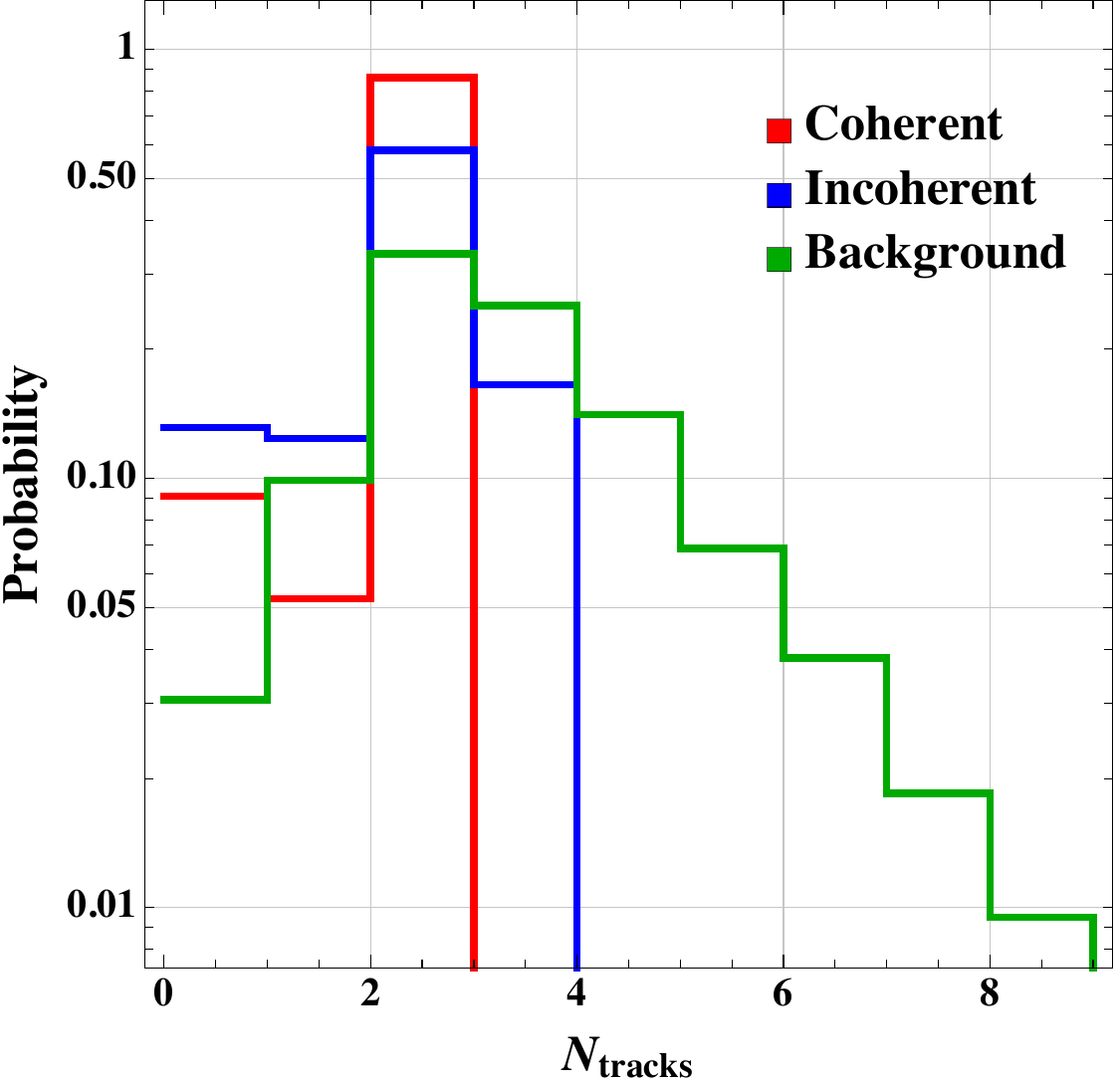} ~~~~
\includegraphics[width=0.45\textwidth]{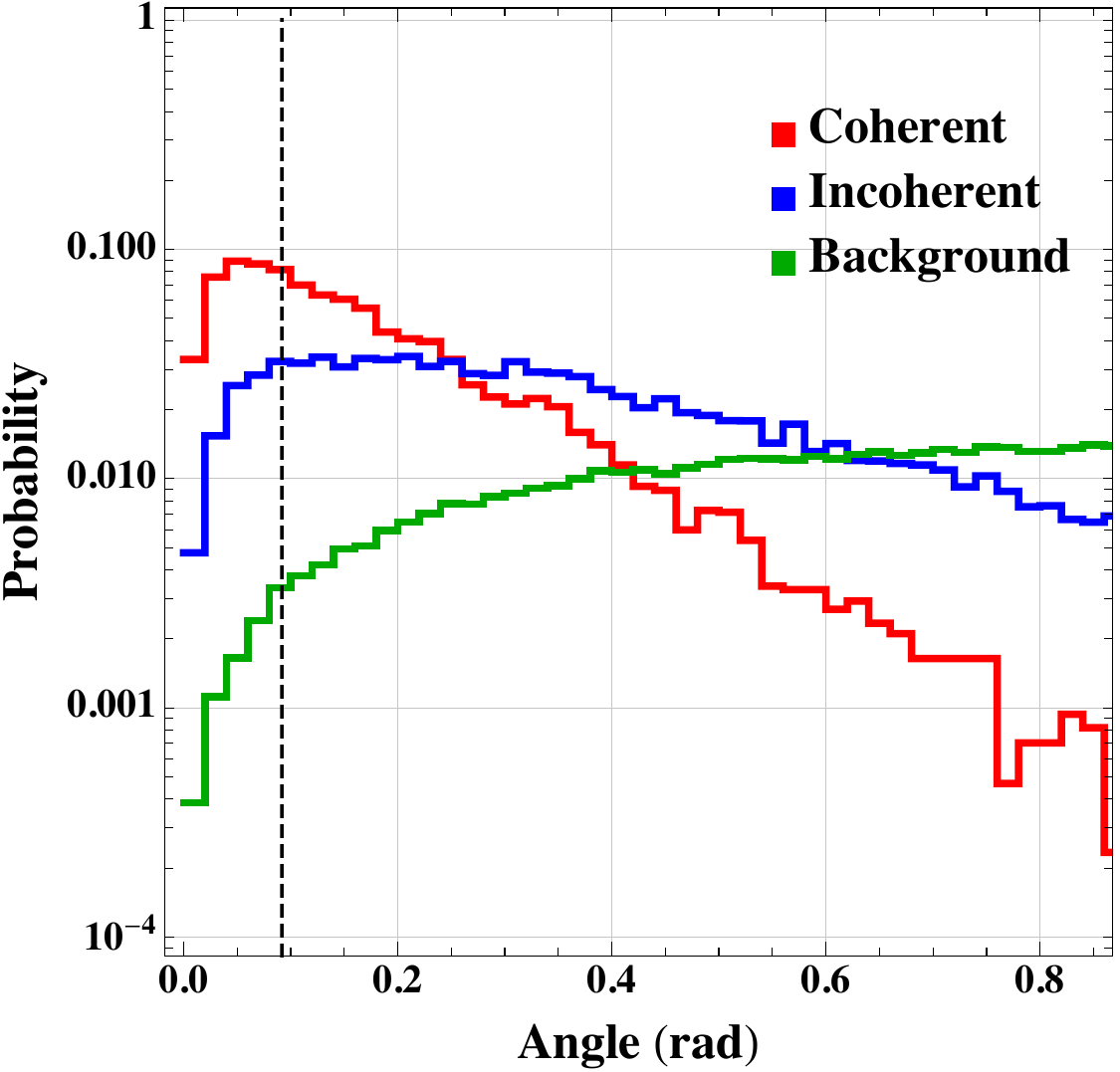} \\[10pt]
\includegraphics[width=0.465\textwidth]{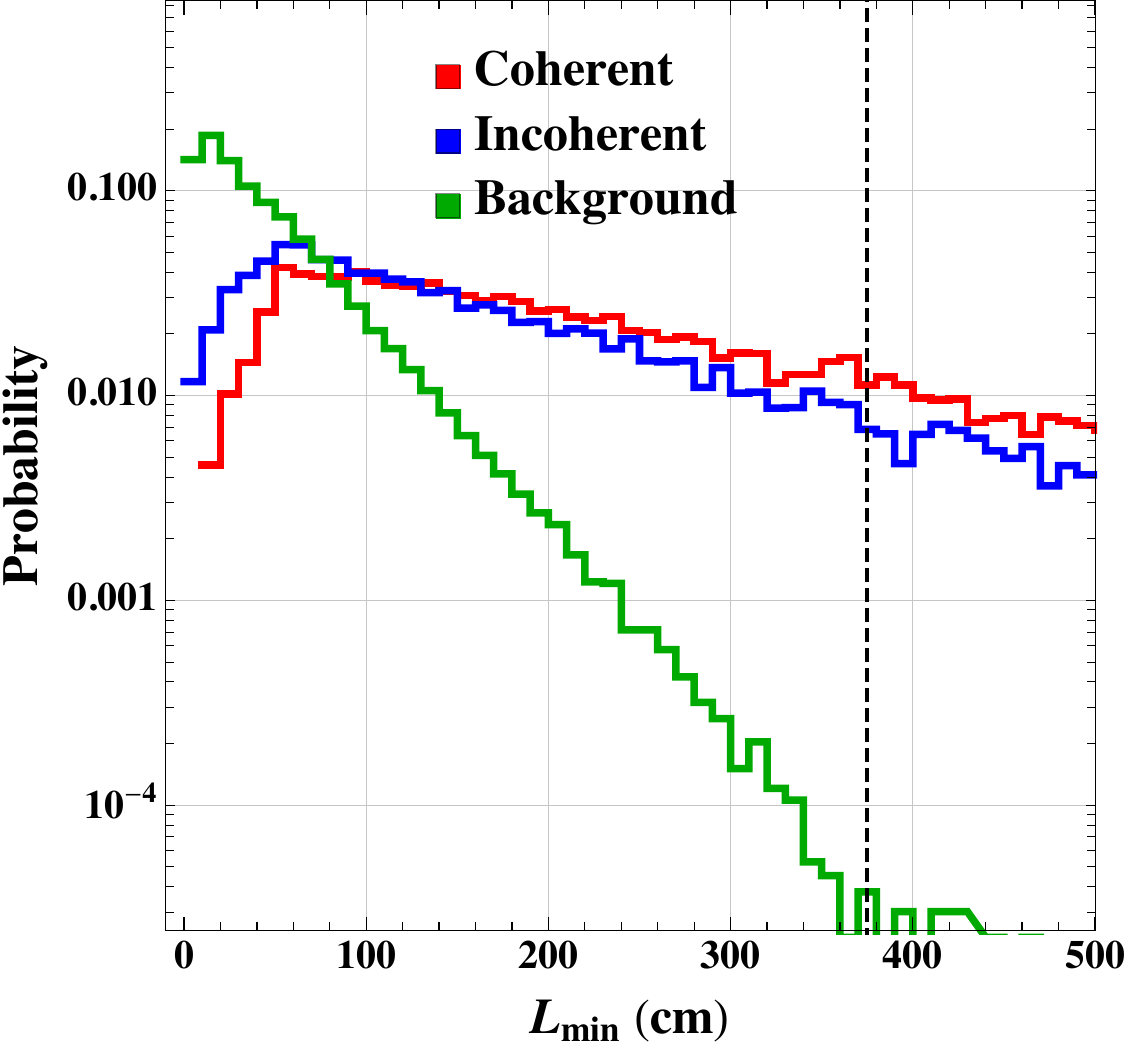} ~
\includegraphics[width=0.46\textwidth]{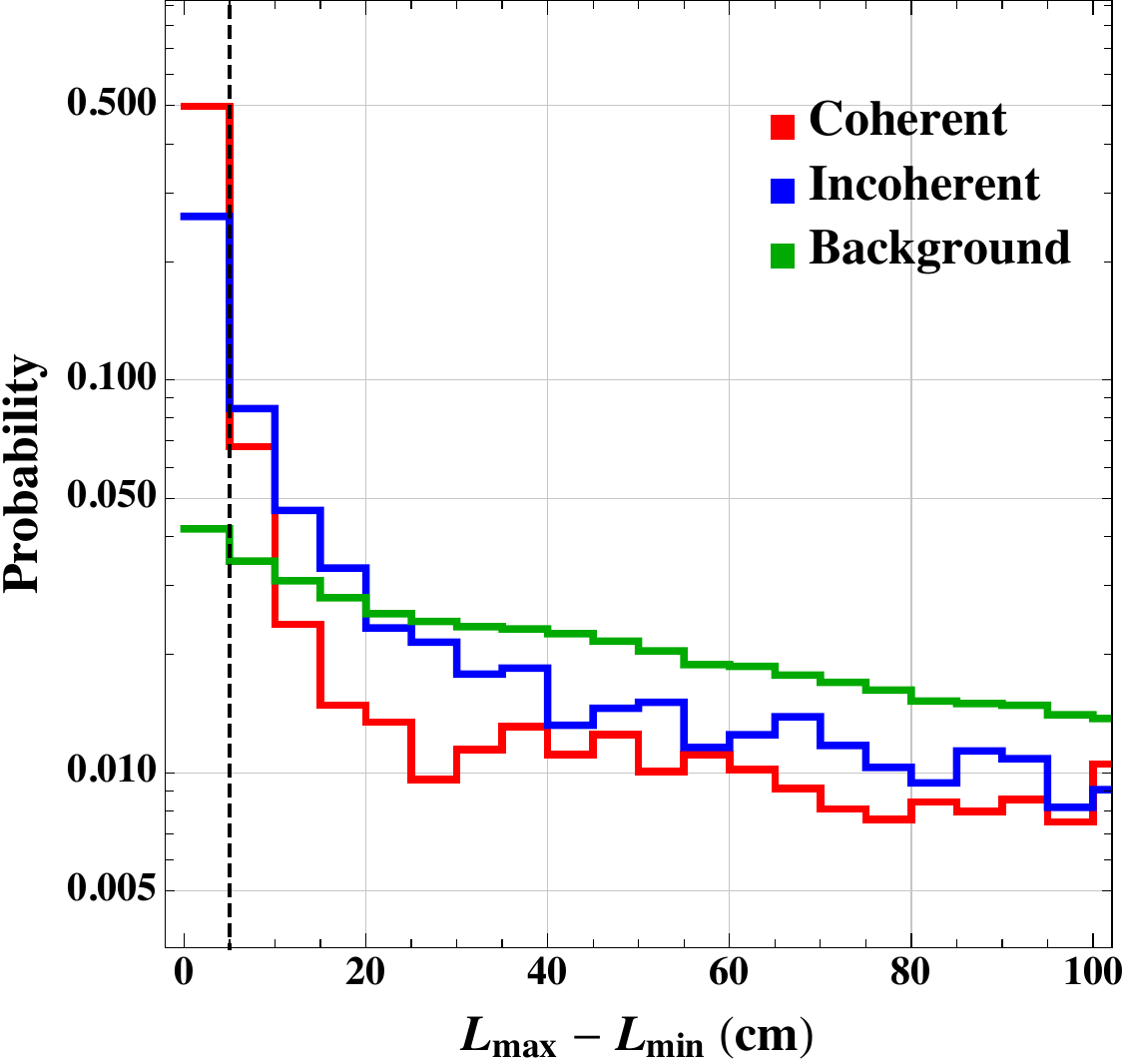} \\[10pt]
\caption{Kinematic distributions for the coherent signal (in red), incoherent signal (in blue), and background (in green) used in our event selection: number of tracks (upper left plot), angle between the two selected tracks (upper right plot), length of the shortest track (lower left plot), and difference in length between the two tracks (lower right plot). For the last three panels, we have only used events containing two and only two tracks. The dashed, black vertical lines indicate the optimized cut used in our analysis (see text for details).} \label{fig:distributionttruth}
\end{figure}

\begin{figure}[tb]
\centering
\includegraphics[width=0.45\textwidth]{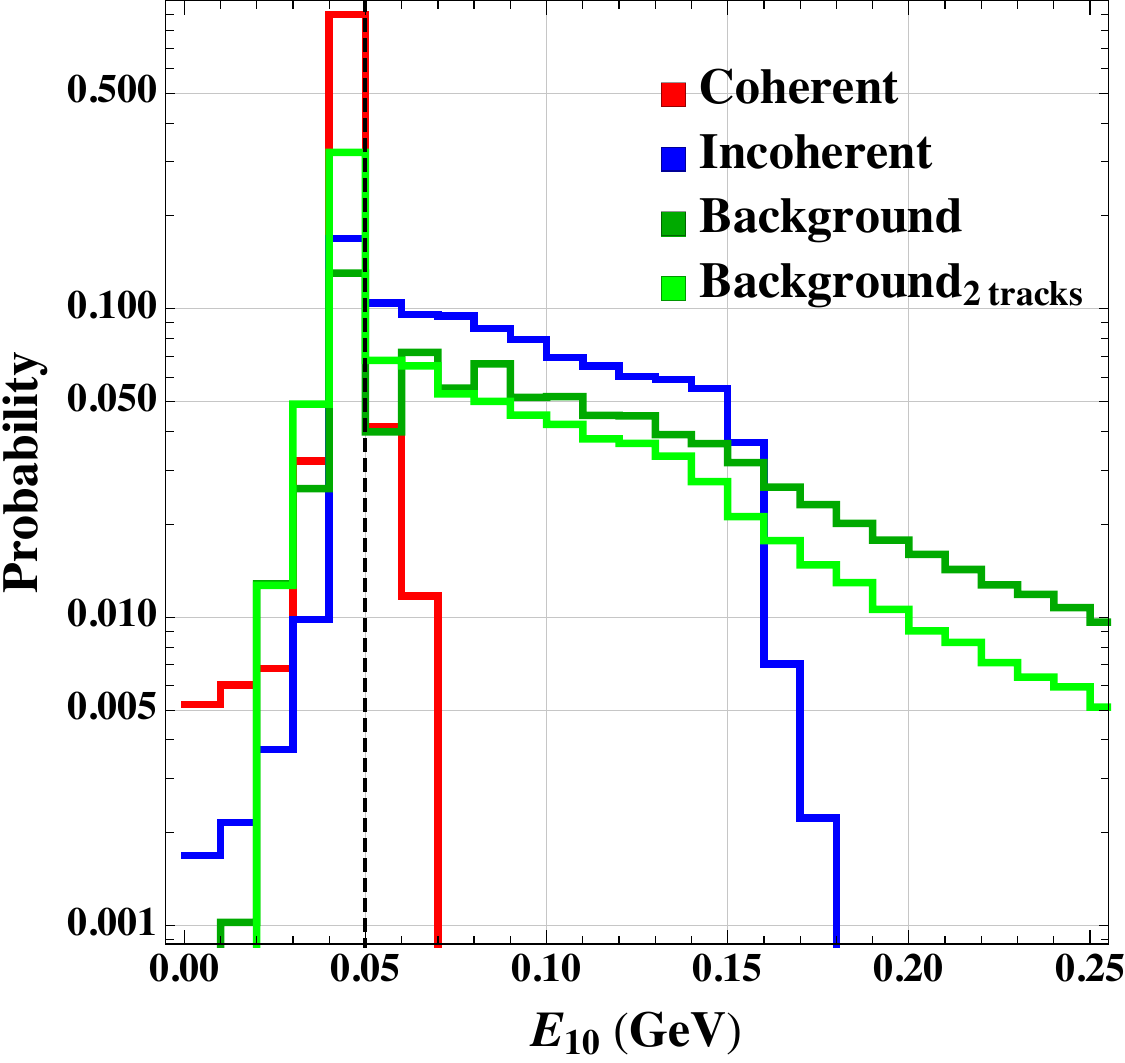}
\caption{Total energy deposited within 10 cm from the neutrino interaction vertex. We show the distribution for the coherent signal (in red), the incoherent signal (in blue), and for background events with and without the requirement of exactly two tracks (lighter and darker green, respectively). The dashed black vertical line indicates the cut used in our analysis.} \label{fig:distributionttruth1}
\end{figure}

Figures~\ref{fig:distributionttruth} and~\ref{fig:distributionttruth1} show the distribution for signal (coherent in red and incoherent in blue) and background (green) events of these kinematic variables. All distributions are area-normalized. Particularly, in the upper left panel of Fig.~\ref{fig:distributionttruth}, we present the distribution for the number of tracks, $N_\mathrm{tracks}$, where we have considered a threshold of 100~MeV in energy deposited in the LAr for the definition of a track. The other panels have been evaluated considering only events that contain two and only two tracks. We consider the distributions for the angle between the two tracks ($angle$, upper right plot), the length of the shortest track ($L_\mathrm{min}$, lower left plot), and the difference in length between the two tracks ($L_\mathrm{max}-L_\mathrm{min}$), lower right plot). Finally, in Fig.~\ref{fig:distributionttruth1} we show the total energy deposited within 10 cm ($E_{10}$) of the neutrino interaction vertex. This includes the sum of the energies deposited by any charged particle (even those that deposit less than 100 MeV and that, therefore, would not be classified as tracks) in a sphere of 10 cm radius around the interaction vertex.

As expected, the background events tend to contain a larger number of tracks than the signal. The other distributions also show a clear discriminating power: the angle between the two tracks is typically much smaller in the signal than in the background. Moreover, the signal tracks (two muons) tend to be longer than tracks in the background events (consisting typically in one muon plus one pion). Finally, the energy deposited in the vicinity of the interaction vertex for the coherent signal events is compatible with the expectation from a pair of minimum ionization tracks, $(\mathrm{d}E/\mathrm{d}x)_\mathrm{mip}\approx2.1~\mathrm{MeV/cm}$. In contrast, both the incoherent signal and the background have, on average, more energy deposited around the vertex due to the hadronic activity generated in the interaction.

\subsection{Expected sensitivity} \label{sec:sensitivity}
The 147-tonne LArTPC at the DUNE near detector will record, in the neutrino-beam mode, close to $3.5\times10^8$ neutrino interactions per year, out of which only a couple hundred events will correspond to the trident process. Our event selection, therefore, has to achieve a background suppression of at least 6 orders of magnitude. 

To do this, we first require events with two and only two tracks, with an angle of at least 0.5 degrees between them to ensure separation of the tracks. This requirement alone is able to suppress the background by a factor of 2, while the signal is almost not affected ($\sim 90\%$ efficiency). On top of this requirement, we optimize the cuts on the other variables shows in Fig.~\ref{fig:distributionttruth}: $angle$, ${L_\mathrm{min}}$, and ${L_\mathrm{max}-L_\mathrm{min}}$. Particularly, we find the values of ${\theta_\mathrm{max}},\,\,{L_M},\,\,{\Delta_L}$ such that the requirements 
\begin{equation}
{angle}<\theta_{\rm{max}},~~{{L_\mathrm{min}}}>L_M,~~{L_\mathrm{max}-L_\mathrm{min}}<\Delta_L,
\end{equation}
produces the largest $S/\sqrt B$ per year. (The following discussion of the optimization of the cuts refers to the trident signal arising from the neutrino mode. The signal acceptance for the antineutrino mode is almost identical.)
We perform a scan on the maximum angle between the two tracks, $\theta_{\rm{max}}$, the minimum length of the shortest track, $L_M$, and the maximum difference between the length of the two tracks, $\Delta_L$. We scan these cuts over a wide range: $\theta_{\rm{max}}\subset [0,0.2]$, $L_M\subset [100,450]\rm{cm}$, $\Delta_L\subset [0,40]\rm{cm}$, in steps of 0.01, 5 cm, and 5 cm, respectively. Each point in this three-dimensional grid will give a number of expected coherent and incoherent signal events, as well as background events. After having obtained this grid, we look for the set of cuts that produces the largest $S/\sqrt B$ per year, after having asked for at least 10 background events in our generated sample. The optimized cuts that we find are given by $\theta_{\rm{max}}=0.09$ ($\sim 5.5$ degree), $L_M=375$ cm, and $\Delta_L=5$ cm. These cuts result in the following number of selected events:
\begin{equation}
S_{\rm{coherent}}\simeq 8.7,~S_{\rm{proton}}\simeq 0.72,~S_{\rm{neutron}}\simeq 0.08,~B \simeq 96,
\end{equation}
per year with $S/\sqrt B\sim 0.9$. 

We now investigate if a cut on the energy deposited in the first 10 cm from the interaction vertex can improve the discrimination of signal vs.\ background (see Fig. \ref{fig:distributionttruth1}). If we require $E_{10}<50$ MeV, the total background (before applying any further cut) is suppressed by a factor of $\sim 5$, while the total signal acceptance is near $70\%$ (this arises from a $\sim 93\%$ acceptance for the coherent signal, a $\sim 7\%$ for the incoherent-proton signal, and a $\sim 85\%$ for the incoherent-neutron signal). 
We find that the cut on $E_{10}$ is correlated with the other cuts we are employing in our analysis. In particular, if we first demand two and only two tracks and, on top of that, $E_{10}<50$ MeV, the suppression of the background due to the $E_{10}$ cut is reduced to a factor of $\sim 3$. This suppression factor is reduced to $\sim 30\%$ when we demand two tracks with a minimum length of 375 cm, as in the previously optimized cuts. Potential systematic uncertainties impacting this vertex periphery cut on the energy deposited have not been included in this analysis. Uncertainties at the level of 10\%, arising mainly from data/simulation discrepancies have been obtained by the MINERvA Collaboration using a similar cut~\cite{Ruterbories:2018gub}. Over the next decade, as the DUNE analysis and simulation framework is developed, such uncertainties should be further reduced. Note that the $E_{10}$ distribution shown in Fig. \ref{fig:distributionttruth1} can
be affected by re-scattering processes that we neglect in our analysis.

We re-run our cut optimization, having asked $E_{10}<50$ MeV. We find that the optimal cuts are only mildly modified to $\theta_{\rm{max}}=0.08$ ($\sim 4.6$ degree), $L_M=340$ cm, and $\Delta_L=4$ cm. These cuts lead to:
\begin{equation}
S_{\rm{coherent}}\simeq 9.8,~S_{\rm{proton}}\simeq 0.18,~S_{\rm{neutron}}\simeq 0.07,~B \simeq 130,
\end{equation}
per year with $S/\sqrt B\sim 0.9$. This shows that the requirement on the vertex activity does not substantially improve the accuracy of the measurement. 

These numbers show that a measurement of the SM di-muon trident production at the 40$\%$ level could be possibly obtained using $\sim 6$ years running in neutrino mode, or, equivalently, $\sim 3$ years running in neutrino mode and $\sim 3$ years running in antineutrino mode.

Given the small expected number of incoherent signal events, $S_\text{proton}$ and $S_\text{neutron}$, a separate measurement of the incoherent cross section appears to be very challenging. Note that our modeling of the kinematics of the nucleon in the incoherent processes might have sizable uncertainties (cf. discussion in section~\ref{sec:resultsAndUncertainties}). However, we do not expect that a more detailed modeling would qualitatively change our conclusions with regards to the incoherent process.


\section{Neutrino tridents and new physics}\label{sec:NP}

Neutrino tridents are induced at the tree level by the electroweak interactions of the SM and thus can probe new interactions among neutrinos and charged leptons of electroweak strength. In the following we discuss the sensitivity of neutrino tridents to heavy new physics parameterized in a model independent way by four fermion interactions (Sec.~\ref{sec:modelind}), and in the context of a new physics model with a light new $Z'$ gauge boson (Sec.~\ref{sec:Zprime}). 

\subsection{Model-independent discussion} \label{sec:modelind}

If the new physics is heavy compared to the relevant momentum transfer in the trident process, its effect is model-independently described by a modification of the effective four fermion interactions introduced in Eq.~(\ref{eq:Heff}).
Focusing on the case of muon-neutrinos interacting with muons, we write
\begin{equation}
g_{\mu\mu\mu\mu}^V = 1 + 4 \sin^2\theta_W + \Delta g_{\mu\mu\mu\mu}^V ~, \qquad g_{\mu\mu\mu\mu}^A = -1 + \Delta g_{\mu\mu\mu\mu}^A ~,
\end{equation}
where $\Delta g_{\mu\mu\mu\mu}^V$ and $\Delta g_{\mu\mu\mu\mu}^A$ parameterize possible new physics contributions to the vector and axial-vector couplings.
Couplings involving other combinations of lepton flavors can be modified analogously. Note, however, that for interactions that involve electrons, very strong constraints can be derived from LEP bounds on electron contact interactions~\cite{Schael:2013ita}.

The modified interactions of the muon-neutrinos with muons alter the cross section of the $\nu_\mu N \to \nu_\mu \mu^+\mu^- N$ trident process. 
We use the existing measurement of the trident cross section by the CCFR experiment~\cite{Mishra:1991bv} and the expected sensitivities at the DUNE near detector discussed in Sec.~\ref{sec:sensitivity}, to put bounds on $\Delta g_{\mu\mu\mu\mu}^V$ and $\Delta g_{\mu\mu\mu\mu}^A$ (see also~\cite{Falkowski:2018dmy}).

Using the neutrino spectrum from the CCFR experiment (see~\cite{McFarland:1995sr}) and the spectrum at the DUNE near detector shown in Fig.~\ref{fig:neutrinoEnergies}, we find the cross sections
\begin{eqnarray}
\sigma_\text{CCFR} &\simeq& (g_{\mu\mu\mu\mu}^V)^2 \times  0.087\,\text{fb}~ + (g_{\mu\mu\mu\mu}^A)^2 \times 0.099\,\text{fb} ~, \\
\sigma_\text{DUNE} &\simeq& (g_{\mu\mu\mu\mu}^V)^2 \times  1.30 \times 10^{-4} \,\text{fb}~ + (g_{\mu\mu\mu\mu}^A)^2 \times 2.00 \times 10^{-4} \,\text{fb} ~,
\end{eqnarray}
where in both cases we only took into account coherent scattering. The CCFR trident measurement put a stringent cut on the hadronic energy at the event vertex region, which we expect to largely eliminate incoherent trident events. Similarly, we anticipate that in a future DUNE measurement incoherent scattering events will be largely removed by cuts on the hadronic activity (see discussion in section~\ref{sec:sensitivity}). 

For the modifications relative to the SM cross sections we find
\begin{eqnarray} \label{eq:ratios}
\frac{\sigma_\text{CCFR}}{\sigma_\text{CCFR}^\text{SM}} &\simeq& \frac{(1+4 \sin^2\theta_W + \Delta g^V_{\mu\mu\mu\mu})^2 + 1.13\, ( 1 -\Delta g^A_{\mu\mu\mu\mu})^2}{(1 + 4 \sin^2\theta_W)^2 + 1.13}~, \\
\frac{\sigma_\text{DUNE}}{\sigma_\text{DUNE}^\text{SM}} &\simeq& \frac{(1+4 \sin^2\theta_W + \Delta g^V_{\mu\mu\mu\mu})^2 + 1.54\, ( 1 - \Delta g^A_{\mu\mu\mu\mu})^2}{(1 + 4 \sin^2\theta_W)^2 + 1.54}~.
\end{eqnarray}

\begin{figure}[tb] \centering
 \includegraphics[width=0.6\textwidth]{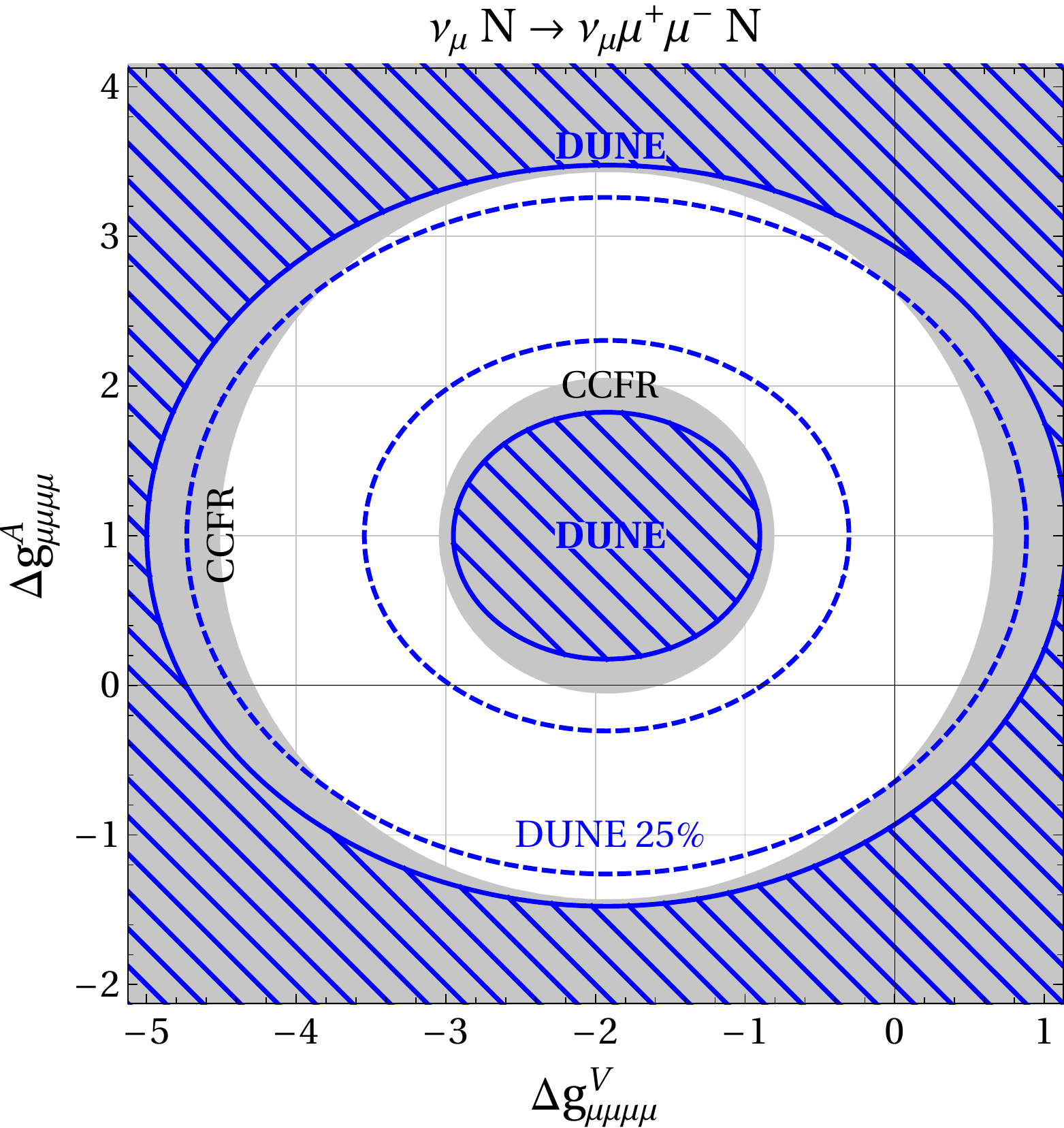}
\caption{95\% CL. sensitivity of a 40\% (blue hashed regions) and a $25\%$ (dashed contours) uncertainty measurement of the $\nu_\mu N \to \nu_\mu \mu^+\mu^- N$ cross section at the DUNE near detector to modifications of the vector and axial-vector couplings of muon-neutrinos to muons. The gray regions are excluded at 95\% CL. by existing measurements of the cross section by the CCFR collaboration. The intersection of the thin black lines indicates the SM point.}
\label{fig:gVgA}
\end{figure}

In Fig.~\ref{fig:gVgA} we show the regions in the $\Delta g^V_{\mu\mu\mu\mu}$ vs. $\Delta g^A_{\mu\mu\mu\mu}$ plane that are excluded by the existing CCFR measurement $\sigma_\text{CCFR} / \sigma_\text{CCFR}^\text{SM} = 0.82 \pm 0.28$~\cite{Mishra:1991bv} at the 95\% C.L. in gray. The currently allowed region corresponds to the white ring including the SM point $\Delta g^V_{\mu\mu\mu\mu} = \Delta g^A_{\mu\mu\mu\mu} = 0$. In the central gray region the new physics interferes destructively with the SM and leads to a too small trident cross section. Outside the white ring, the trident cross section is significantly larger than observed. The result of our baseline analysis (corresponding to an expected measurement with 40\% uncertainty) does not extend the sensitivity into parameter space that is unconstrained by the CCFR measurement. However, it is likely that the use of a magnetized spectrometer, as it is being considered for the DUNE ND, able to identify the charge signal of the trident final state, along with more sophisticated deep-learning based event selection, will significantly improve separation between neutrino trident interactions and backgrounds. Therefore, we also present the region that could be probed by a 25\% measurement of the neutrino trident cross section at DUNE, which would extend the coverage of new physics parameter space substantially.

\subsection{$Z'$ model based on gauged $L_\mu - L_\tau$} \label{sec:Zprime}

A class of example models that modify the trident cross section are models that contain an additional neutral gauge boson, $Z'$, that couples to neutrinos and charged leptons. A consistent way of introducing such a $Z'$ is to gauge an anomaly free global symmetry of the SM. Of particular interest is the $Z'$ that is based on gauging the difference between muon-number and tau-number, $L_\mu - L_\tau$~\cite{He:1990pn,He:1991qd}. Such a $Z'$ is relatively weakly constrained and can for example address the longstanding discrepancy between SM prediction and measurement of the anomalous magnetic moment of the muon, $(g-2)_\mu$~\cite{Baek:2001kca,Harigaya:2013twa}. The $L_\mu - L_\tau$ $Z'$ has also been used in models to explain $B$ physics anomalies~\cite{Altmannshofer:2014cfa} and as a portal to dark matter~\cite{Baek:2008nz,Altmannshofer:2016jzy}. The $\nu_\mu N \to \nu_\mu \mu^+\mu^- N$ trident process has been identified as important probe of gauged $L_\mu - L_\tau$ models over a broad range of $Z^\prime$ masses~\cite{Altmannshofer:2014cfa,Altmannshofer:2014pba}.

The interactions of the $Z'$ with leptons and neutrinos are given by
\begin{equation}\label{LmuLtau}
 \mathcal{L}_{L_\mu-L_\tau} = g' Z'_\alpha \Big[ (\bar\mu \gamma^\alpha \mu) - (\bar\tau \gamma^\alpha \tau)  + (\bar\nu_\mu \gamma^\alpha P_L \nu_\mu) - (\bar\nu_\tau \gamma^\alpha P_L \nu_\tau) \Big]~,
\end{equation}
where $g'$ is the $L_\mu-L_\tau$ gauge coupling. Note that the $Z'$ couples purely vectorially to muons and taus. If the $Z'$ is heavy when compared to the momentum exchanged in the process, it can be integrated out, and its effect on the $\nu_\mu N \to \nu_\mu \mu^+\mu^- N$ process is described by the effective couplings
\begin{equation}
 \Delta g^V_{\mu\mu\mu\mu} = (g')^2 \frac{2 v^2}{m_{Z'}^2} ~,\qquad \Delta g^A_{\mu\mu\mu\mu} = 0 ~,
\end{equation}
where $m_{Z'}$ is the $Z'$ mass and $v \simeq 246$~GeV is the electroweak breaking vacuum expectation value.
Using the expression for the cross section in~(\ref{eq:ratios}) we find the following bound from the existing CCFR measurement
\begin{equation} \label{eq:gprime}
 g' \lesssim  0.2 \times \left(\frac{m_{Z'}}{100~\text{GeV}}\right) \qquad \text{for} ~ m_{Z'} \gtrsim \text{few~GeV} ~.
\end{equation}
The bound is applicable as long as the $Z^\prime$ mass is heavier than the average momentum transfer in the trident reaction at CCFR, which -- given the neutrino energy spectrum at CCFR -- is around a few GeV.  
For lower $m_{Z'}$, the $Z'$ propagator is saturated by the momentum transfer and the CCFR bound on $g'$ improves only logarithmically.
A measurement of the trident process at the DUNE near detector has the potential to considerably improve the sensitivity for low-mass $Z'$ bosons. Because of the much lower energy of the neutrino beam compared to CCFR, also the momentum transfer is much smaller and the scaling in eq.~(\ref{eq:gprime}) extends to smaller $Z'$ masses.

\begin{figure}[!tb] \centering
\includegraphics[width=0.9\textwidth]{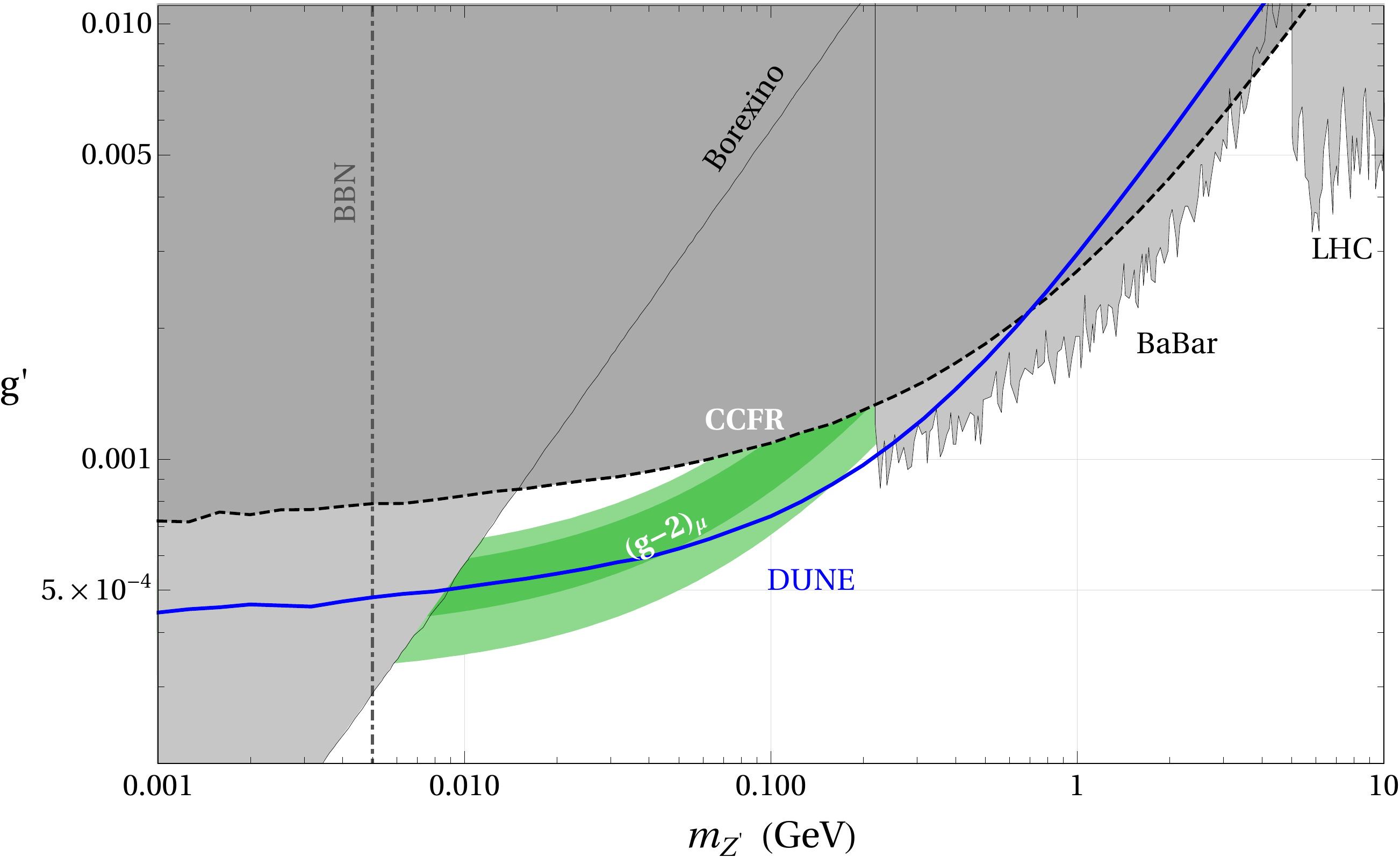}
\caption{Existing constraints and DUNE sensitivity in the $L_\mu - L_\tau$ parameter space. Shown in green is the region where the $(g-2)_\mu$ anomaly can be explained at the $1\sigma$ and $2\sigma$ level. The parameter regions already probed by existing constraints are shaded in gray and correspond to a CMS search for $pp \to \mu^+\mu^- Z' \to \mu^+\mu^-\mu^+\mu^-$~\cite{Sirunyan:2018nnz} (``LHC''), a BaBar search for $e^+e^- \to \mu^+\mu^- Z' \to \mu^+\mu^-\mu^+\mu^-$~\cite{TheBABAR:2016rlg} (``BaBar''), precision measurements of $Z \to \ell^+ \ell^-$ and $Z \to \nu\bar\nu$ couplings~\cite{ALEPH:2005ab,Altmannshofer:2014cfa} (``LEP''), a previous measurement of the trident cross section~\cite{Mishra:1991bv,Altmannshofer:2014pba} (``CCFR''), a measurement of the scattering rate of solar neutrinos on electrons~\cite{Bellini:2011rx,Harnik:2012ni,Agostini:2017ixy} (``Borexino''), and bounds from Big Bang Nucleosynthesis~\cite{Ahlgren:2013wba,Kamada:2015era,Escudero:2019gzq} (``BBN''). The DUNE sensitivity shown by the solid blue line assumes 6.5 year running in neutrino mode, leading to a measurement of the trident cross section with $40\%$ precision.}
\label{fig:LmuLtau}
\end{figure}

In Fig.~\ref{fig:LmuLtau} we show the existing CCFR constraint on the model parameter space in the $m_{Z'}$ vs. $g'$ plane and compare it to the region of parameter space where the anomaly in $(g-2)_\mu = 2 a_\mu$ can be explained. The green region shows the $1\sigma$ and $2\sigma$ preferred parameter space corresponding to a shift $\Delta a_\mu = a_\mu^\text{exp}-a_\mu^\text{SM} = (2.71 \pm 0.73) \times 10^{-9}$~\cite{Keshavarzi:2018mgv} (see also \cite{Davier:2017zfy}).
In the figure, we also show the constraints from LHC searches for the $Z'$ in the $pp \to \mu^+\mu^- Z' \to \mu^+\mu^-\mu^+\mu^-$ process~\cite{Sirunyan:2018nnz,Altmannshofer:2014pba} (see also~\cite{Chun:2018ibr}), direct searches for the $Z'$ at BaBar using the $e^+e^- \to \mu^+\mu^- Z' \to \mu^+\mu^-\mu^+\mu^-$ process~\cite{TheBABAR:2016rlg}, and constraints from LEP precision measurements of leptonic $Z$ couplings~\cite{ALEPH:2005ab,Altmannshofer:2014cfa}.  
Also a Borexino bound on non-standard contributions to neutrino-electron scattering~\cite{Harnik:2012ni,Bellini:2011rx,Agostini:2017ixy} has been used to constrain the $L_\mu - L_\tau$ gauge boson~\cite{Kamada:2015era,Araki:2015mya,Kamada:2018zxi}. Our version of this constraint (see appendix~\ref{app:borexino}) is also shown.
For very light $Z'$ masses of $O$(few MeV) and below, strong constraints from measurements of the effective number of relativistic degrees of freedom during Big Bang Nucleosynthesis (BBN) apply~\cite{Ahlgren:2013wba,Kamada:2015era,Escudero:2019gzq} (see the vertical dot-dashed line in the figure). For $m_Z^\prime \simeq 10$~MeV, the tension in the Hubble parameter $H_0$ can be ameliorated~\cite{Escudero:2019gzq}.
Taking into account all relevant constraints, the region of parameter space left that may explain $(g-2)_\mu$ lies below the di-muon threshold $m_{Z'}\lesssim 210$~MeV.

A measurement of the $\nu_\mu N \to \nu_\mu \mu^+\mu^- N$ cross section at the SM value with $40\%$ uncertainty at the DUNE near detector is sensitive to the region delimited by the blue contour. We find that the parameter space that is motivated by $(g-2)_\mu$ could be covered in its majority.

Other proposals to cover the remaining region of parameter space favored by $(g-2)_\mu$ include LHC searches for $\mu^+\mu^- + E_T\!\!\!\!\!\!/$\,\,~~\cite{Elahi:2015vzh}, searches for  $\gamma + E\!\!\!/$\, at Belle II~\cite{Araki:2017wyg}, muon fixed-target experiments~\cite{Gninenko:2014pea,Kahn:2018cqs}, high-intensity electron fixed-target experiments~\cite{Gninenko:2018tlp}, or searches for $Z + E\!\!\!/$\, at future electron-positron colliders~\cite{Nomura:2018yej}.

\section{Conclusions} \label{sec:conclusions}

The production of a pair of charged leptons through the scattering of a neutrino on a heavy nucleus (i.e.\ neutrino trident production) is a powerful probe of new physics in the leptonic sector. In this paper we have studied the sensitivity to this process of the planned DUNE near detector.

In the SM, neutrino trident production proceeds via the weak interaction, and thus the cross section can be computed with good accuracy. Here, we have provide SM predictions for the cross sections and the expected rates at the DUNE near detector for a variety $\nu_\mu$ and $\overline{\nu}_\mu$ of neutrino and antineutrino-induced trident processes: $\overset{(-)}{\nu}_\mu \to \overset{(-)}{\nu}_\mu \mu^+\mu^-$, $\overset{(-)}{\nu}_\mu \to \overset{(-)}{\nu}_\mu e^+e^-$, and $\overset{(-)}{\nu}_\mu \to \overset{(-)}{\nu}_e e^\pm \mu^\mp$.
We estimate that the uncertainties of our predictions for the dominant coherent scattering process are approximately $6\%$, mainly due to higher order electroweak corrections. 
Sub-dominant contributions from incoherent scattering have larger uncertainties due to nuclear modeling.

We find that at the DUNE near detector, one can expect $\sim 240$ $\nu_\mu \to \nu_\mu \mu^+\mu^-$ events per year, $\sim 440$ $\nu_\mu \to \nu_\mu e^+e^-$ events per year, and $\sim 1600$ $\nu_\mu \to \nu_e e^+\mu^-$ events per year. This implies favorable conditions for performing precise measurements of the cross sections of such processes. 

In this paper, we performed a state-of-the-art analysis for the future sensitivity of DUNE to muon neutrino tridents using a Geant4-based simulation of the DUNE near detector liquid argon TPC. Thanks to the very distinctive kinematical features of the signal, if compared to the muon inclusive production background (two long tracks with a relatively small opening angle and a small energy deposited around the neutrino vertex interaction), the background rate can be reduced by $\sim 7$ order of magnitude, reaching levels comparable to the signal. We expect to be able to observe $\mathcal O(10)$ signal event/year and $\mathcal O(20)$ background/year. The main source of background arises from pion-muon production, where both the pion and the muon produce long tracks, and originate from the first part of the liquid argon. A further suppression of the background might be obtained via the magnetized spectrometer, whose sampling calorimeter should improve the separation between muons and pions.


We find that the $\nu_\mu \to \nu_\mu \mu^+\mu^-$ trident cross section can be measured with good precision at the DUNE near detector. Taking into account approximately three years running each in neutrino and antineutrino mode, we anticipate a measurement with an accuracy of $\sim 40\%$. This is comparable to the accuracy of the measurements by the CCFR and CHARM II collaborations, see eq.~(\ref{eq:trident_exp}). Note, however, that the much lower energy of the neutrino beam at DUNE leads to an enhanced sensitivity to light new physics. Moreover, it is likely that the use of the magnetized spectrometer, along with more sophisticated deep-learning based event selection, will significantly improve the accuracy at DUNE.

We also analyzed the impact of such a measurement on physics beyond the SM, both model independently and in a benchmark $Z^\prime$ model. We find that a measurement at DUNE could significantly extend the coverage of new physics parameter space compared to the existing trident measurement from the CCFR and CHARM II experiments. This is particularly the case for light new physics.  
As a benchmark new physics model we considered an extension of the SM by a new $Z^\prime$ gauge boson that is based on gauging the difference of muon-number and tau-number, $L_\mu - L_\tau$. We provide a summary of existing constraints on the $Z^\prime$ parameter space in Fig.~\ref{fig:LmuLtau}. Interestingly enough, there is viable parameter space where the $Z^\prime$ can explain the long-standing discrepancy in the anomalous magnetic moment of the muon, $(g-2)_\mu$. We find that the parameter space that is motivated by $(g-2)_\mu$ could be largely covered by a measurement of the $\nu_\mu \to \nu_\mu \mu^+\mu^-$ trident cross section at DUNE.

\section*{Acknowledgements}

We thank Chris Ontko for the collaboration in the early stages of this paper. We thank the DUNE Collaboration for reviewing this manuscript and providing computing resources for the simulation of neutrino interactions in the DUNE near detector. The research of WA is supported by the National Science Foundation under Grant No.\ PHY-1912719. SG is supported by a National Science Foundation CAREER Grant No.\ PHY-1915852. The work of WA and SG was in part performed at the Aspen Center for Physics, which is supported by National Science Foundation Grant PHY-1607611. The work of AS and MW was supported by the Office of High Energy Physics at the Department of Energy through grant DE-SC011784 to the University of Cincinnati.

\begin{appendix}
\section{Nuclear form factors} \label{app:ff}

\begin{figure}[tb] 
 \includegraphics[width=0.48\textwidth]{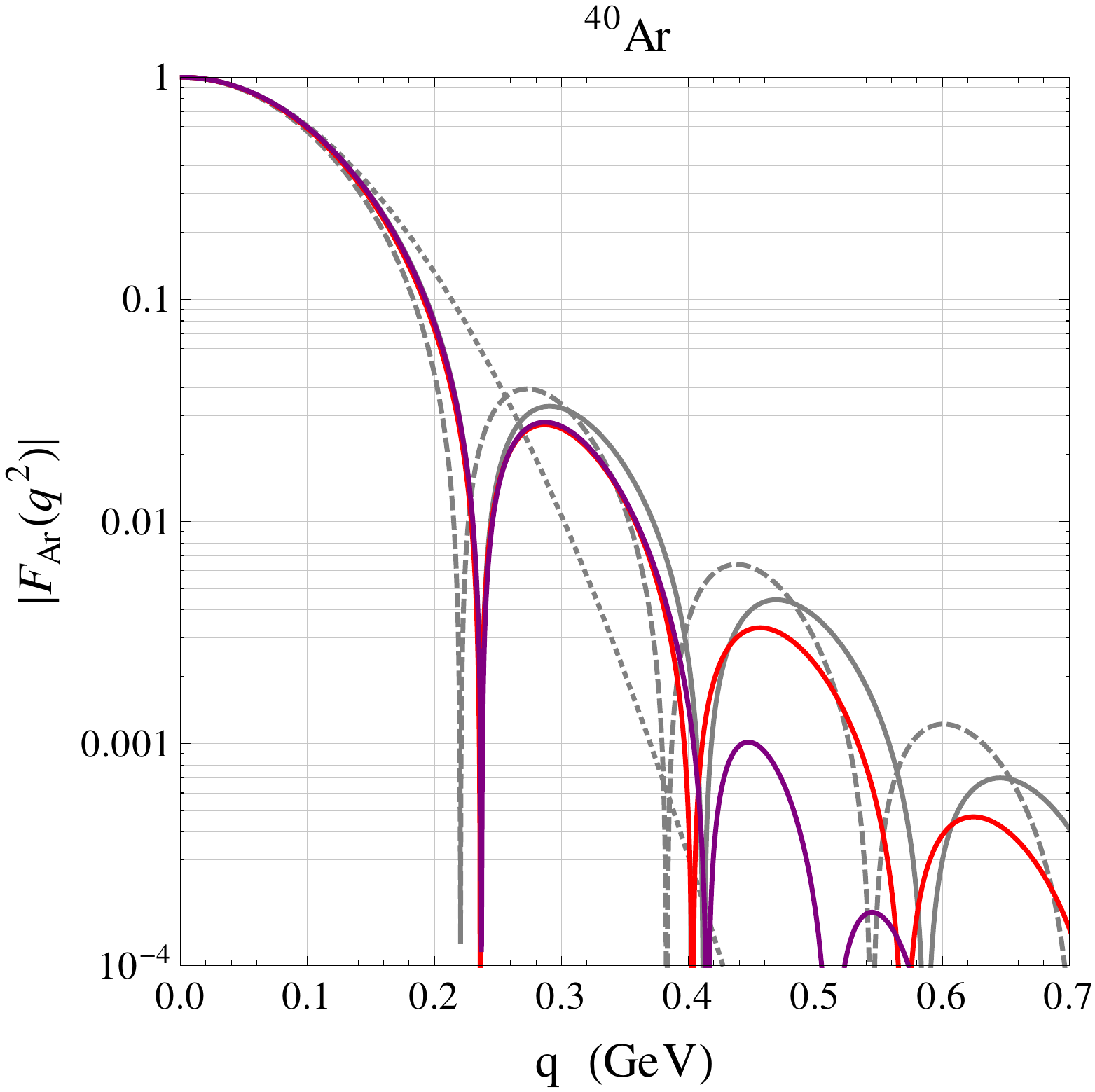} ~~~
 \includegraphics[width=0.48\textwidth]{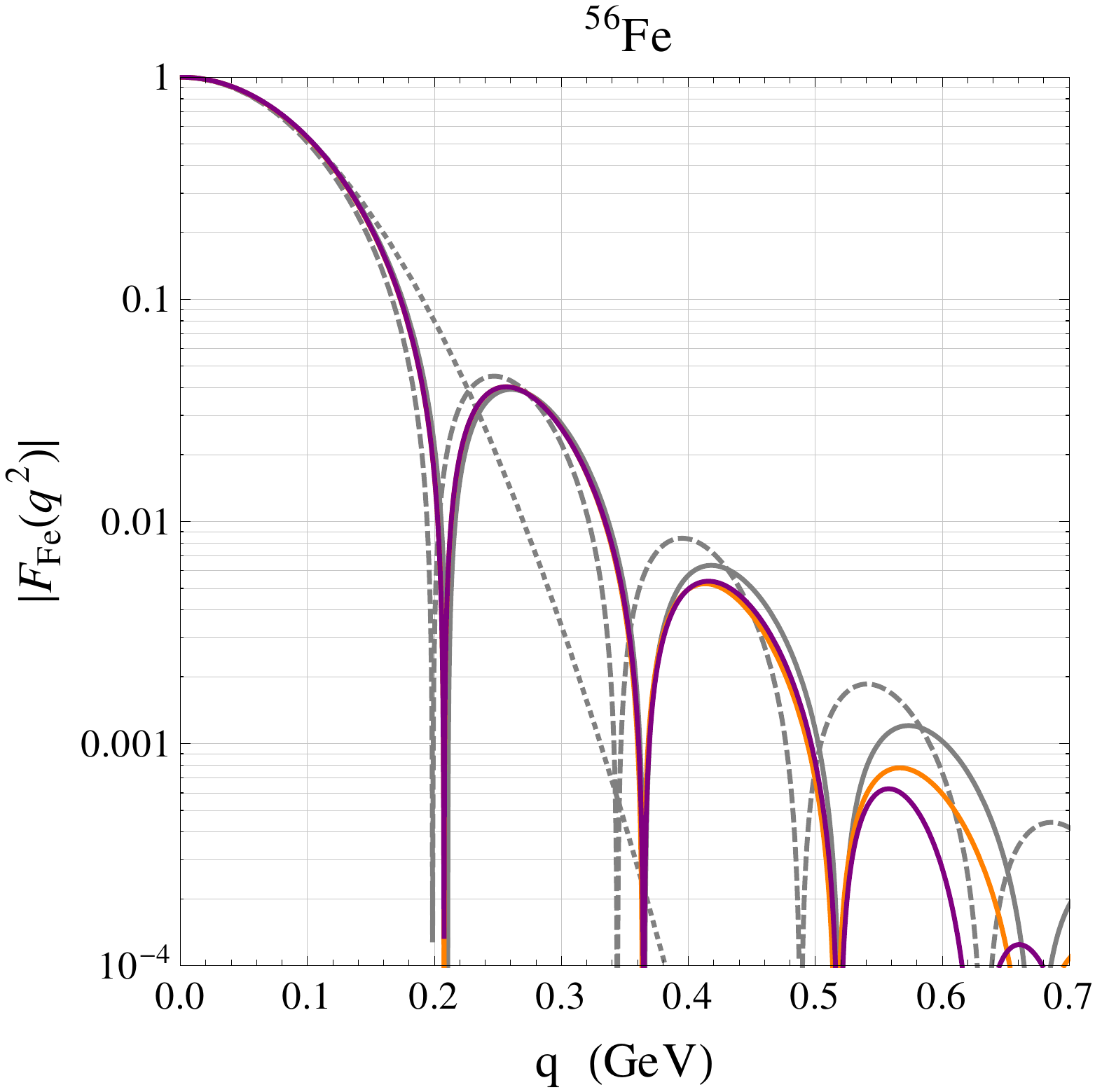} 
\caption{Electric form factors, $F_N(q^2)$, of argon (left panel) and iron (right panel) based on various parameterizations of the nuclear charge density distributions. See text for details.}
\label{fig:ff}
\end{figure}

In our predictions for the coherent neutrino trident process we use electric form factors based on nuclear charge density distributions that have been fitted to elastic electron scattering data~\cite{DeJager:1987qc}. The form factors are expressed as
\begin{equation}
 F_N(q^2) = \int dr ~r^2 ~\frac{\sin(qr)}{qr} \rho_N(r) ~,
\end{equation}
where $q = \sqrt{q^2}$ and $\rho_N$ is a spherically symmetric charge density distribution of the nucleus~$N$, normalized as $\int dr ~r^2 \rho_N(r) = 1$, such that $F_N(0) = 1$. 
The charge distributions $\rho_N$ can be parameterized in various different ways. 
In Fig.~\ref{fig:ff} we compare the form factors for argon and iron that we obtain using various parameterizations that are available from~\cite{DeJager:1987qc}. 
We show the form factors based on charge densities parameterized by a Fourier-Bessel series expansion in purple. Form factors based on the three parameter Fermi charge distribution and the three parameter Gaussian charge distribution are shown in red and orange, respectively.

We also compare these form factors with other phenomenological parameterizations which are much less precise.
In particular we consider a two parameter charge density distribution
\begin{equation}
 \rho_N(r) = \frac{\mathcal N}{1 + \exp\left\{ (r - r_0)/\sigma \right\} } ~,
\end{equation}
with $r_0 = 1.18~\text{fm} \times A^\frac{1}{3} - 0.48$~fm and $\sigma = 0.55$~fm~\cite{Lovseth:1971vv} (shown in solid gray in the plots of Fig.~\ref{fig:ff}) and $r_0 = 1.126~\text{fm} \times A^\frac{1}{3}$ and $\sigma = 0.523$~fm~\cite{Magill:2016hgc,Ballett:2018uuc} (shown in dashed gray), where $A$ is the mass number of the nucleus. Finally, in dotted gray, we show a simple exponential form factor~\cite{Brown:1973ih}
\begin{equation}
 F_N(q^2) = \exp\left\{ - \frac{a^2 q^2}{10} \right\} ~,~~ \text{with} ~ a = 1.3~\text{fm} \times A^\frac{1}{3} ~.
\end{equation}
This is the form factor that was used in~\cite{Altmannshofer:2014pba}.

We observe that the form factors that are based on the fitted nuclear charge distributions from~\cite{DeJager:1987qc} (purple, orange, and red lines in the figure) agree very well over a broad range of relevant momentum transfer. Our predictions for the trident cross sections differ by less than 1\% using these parameterizations. 
As default for our numerical calculations we choose the Fourier-Bessel series expansion for argon and iron
\begin{equation}\label{eq:A4}
 \rho_N(r) = \begin{cases} \mathcal{N} \sum_n a_N^{(n)}  j_0(n \pi r / R_N)  & \text{for} ~~r < R_N ~, \\ 0 & \text{for} ~~r > R_N ~, \end{cases}
\end{equation}
where $j_0$ is the spherical Bessel function of 0-th order. (These correspond to the purple lines in Fig.~\ref{fig:ff}).
The coefficients $a_N^{(n)}$ and $R_N$ are given in Table~\ref{tab:ff}. 

\begin{table}[tb] 
\begin{center}
\begin{tabular}{ccc|ccc}
\hline\hline
~~~~~~~~~~~~ & ~~~~~~~~argon~~~~~~~~ & ~~~~~~~~iron~~~~~~~~ & ~~~~~~~~~~~~ & ~~~~~~~~argon~~~~~~~~ & ~~~~~~~~iron~~~~~~~~ \\
\hline
$R$		& $9$~fm & $9$~fm & $a^{(9)}$	 & $0.00011971$ 	 & $-0.00018146$ \\
$a^{(1)}$	&  $0.030451$  & $0.042018$ & $a^{(10)}$	 & $-0.000019801$  	 & $0.00037261$ \\
$a^{(2)}$	& $0.055337$ 	 & $0.062337$ & $a^{(11)}$	& $-0.0000043204$ 	 & $-0.00023296$ \\
$a^{(3)}$	& $0.020203$ 	 & $0.00023995$ & $a^{(12)}$	& $0.0000061205$ 	 & $0.00011494$ \\
$a^{(4)}$	& $-0.016765$ 	 & $-0.032776$  & $a^{(13)}$	& $-0.0000037803$ 	 & $-0.000050596$ \\
$a^{(5)}$	& $-0.013578$ 	 & $-0.0079941$ & $a^{(14)}$	& $0.0000018001$ 	 & $0.000020652$ \\
$a^{(6)}$	& $-0.000043204$ 	 & $0.010844$ & $a^{(15)}$	& $-0.00000077407$ 	 & $-0.0000079428$ \\
$a^{(7)}$	& $0.00091988$ 	 & $0.0049123$ & $a^{(16)}$	& -- 	 & $0.0000028986$ \\
$a^{(8)}$	& $-0.00041205$ 	 & $-0.0022144$ & $a^{(17)}$	& -- 	 & $-0.0000010075$ \\
\hline\hline
\end{tabular}
\end{center}
\caption{Parameters for the Fourier-Bessel series expansion of the charge density distribution of argon and iron from~\cite{DeJager:1987qc}, see Eq. (\ref{eq:A4})}.
\label{tab:ff}
\end{table}

The trident cross sections that we obtain with the two parameter form factors used in~\cite{Lovseth:1971vv} differ at most by few \% from the results using our default form factors.
We find that the form factors used in~\cite{Magill:2016hgc,Ballett:2018uuc} (dashed gray in the figure) tend to be somewhat smaller than the others and underestimate the trident cross sections by approximately $10\%$ for both argon and iron. The simple exponential form factors used in~\cite{Altmannshofer:2014pba} (dotted gray in the figure) give cross sections that agree reasonably well in the case of iron, but tend to overestimate the cross sections for argon by $5\%-10\%$.

\section{Nucleon form factors} \label{app:ff2}

We use proton and neutron form factors from~\cite{Alberico:2008sz} (see also~\cite{Ye:2017gyb} for a recent re-evaluation of nucleon form factors). The form factors of the proton were obtained from fits to measurements of the electron-proton elastic scattering cross section and polarization transfer measurements. The form factors of the neutron were obtained from fits to electron-nucleus (mainly deuterium and $^3$He) scattering data.
The following parameterizations are used for the electric form factor $G_E^p$ and magnetic form factor $G_M^p$ of the proton
\begin{eqnarray} \label{eq:GEp}
 G_E^p(q^2) &=& \frac{1 + a_p^E \tau}{ 1 + b_{p,1}^E \tau + b_{p,2}^E \tau^2 + b_{p,3}^E \tau^3}~, \\
 \frac{G_M^p(q^2)}{\mu_p} &=& \frac{1 + a_p^M \tau}{ 1 + b_{p,1}^M \tau + b_{p,2}^M \tau^2 + b_{p,3}^M \tau^3}~,
\end{eqnarray}
where $\tau = q^2/(4 m_p^2)$ and the magnetic moment of the proton is $\mu_p \simeq 2.793$. An analogous parameterization is used for the magnetic form factor of the neutron 
\begin{equation}\label{eq:GEmn}
 \frac{G_M^n(q^2)}{\mu_n} = \frac{1 + a_n^M \tau}{ 1 + b_{n,1}^M \tau + b_{n,2}^M \tau^2 + b_{n,3}^M \tau^3}~,
\end{equation}
where $\tau = q^2/(4 m_n^2)$ and the magnetic moment of the neutron is $\mu_n \simeq -1.913$. 
The $a$ and $b$ parameters are collected in Table~\ref{tab:ff2}. 

\begin{table}[tb] 
\begin{center}
\begin{tabular}{lcc|lc}
\hline\hline
& ~~~~~proton~~~~~ & ~~~~~neutron~~~~~ & & ~~~~~proton~~~~~ \\
\hline
~~$a^M$		&  1.09		& 8.28  & ~~~~$a^E$	&  -0.19\\
~~$b_1^M$	&  12.31	& 21.3  & ~~~~$b_1^E$	&  11.12\\
~~$b_2^M$	&  25.57	& 77  	& ~~~~$b_2^E$	&  15.16\\
~~$b_3^M$	&  30.61	& 238   & ~~~~$b_3^E$	&  21.25\\
\hline\hline
\end{tabular}
\end{center}
\caption{Parameters for the electric and magnetic form factors of the proton and neutron from~\cite{Alberico:2008sz}, see Eqs.~(\ref{eq:GEp})~-~(\ref{eq:GEmn}).
}
\label{tab:ff2}
\end{table}

Finally, the electric form factor of the neutron is parametrized in the following way
\begin{equation} \label{eq:GEn}
 G_E^n(q^2) = \frac{A \tau}{ 1 + B \tau} G_D(q^2)~,
\end{equation}
where $A = 1.68$, $B = 3.63$, and the standard dipole form factor is $G_D(q^2) = (1 + q^2 / m_V^2)^{-2}$, with $m_V^2 = 0.71$~GeV$^2$.

\begin{figure}[tb] 
 \includegraphics[width=0.48\textwidth]{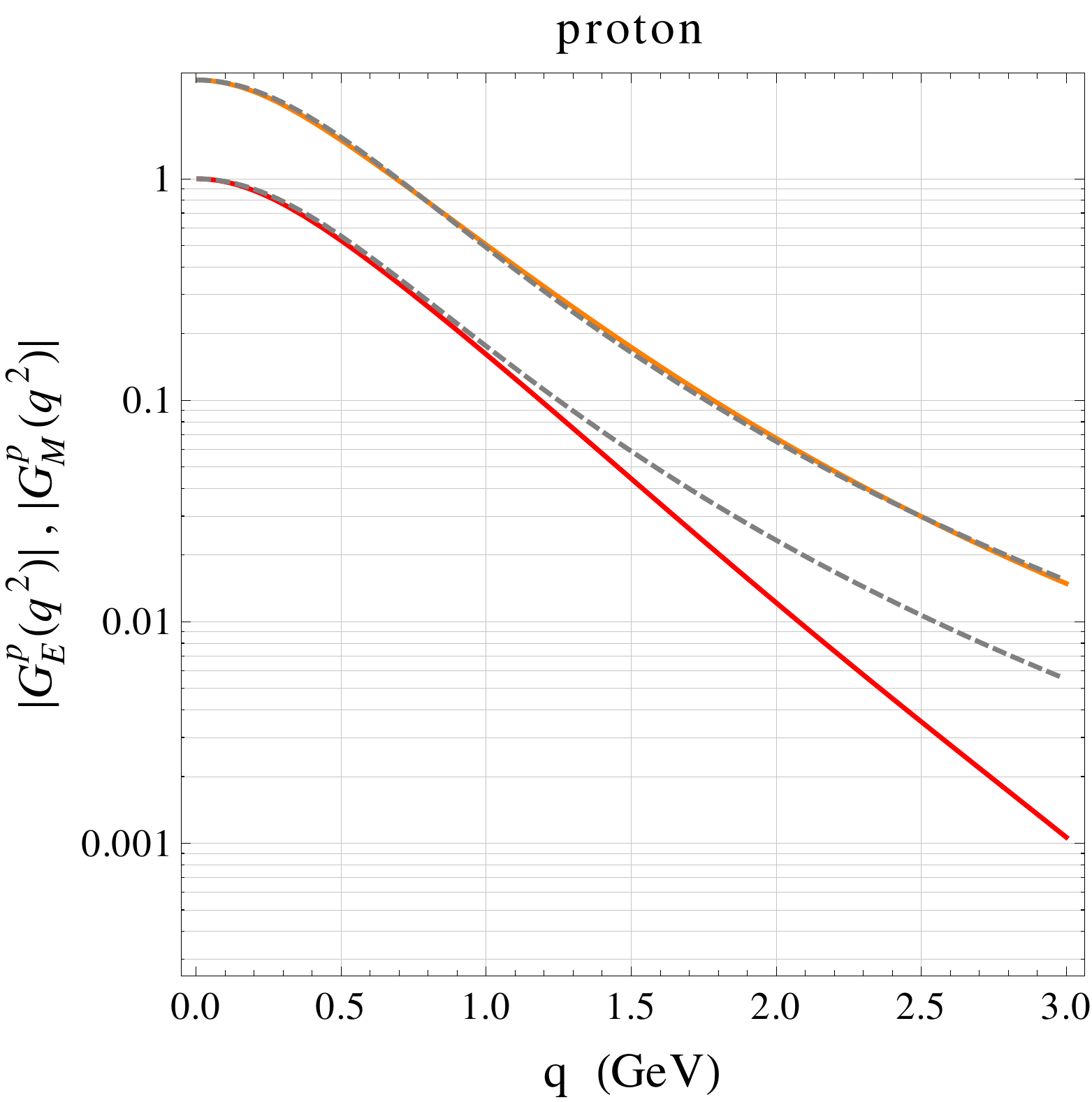} ~~~
 \includegraphics[width=0.48\textwidth]{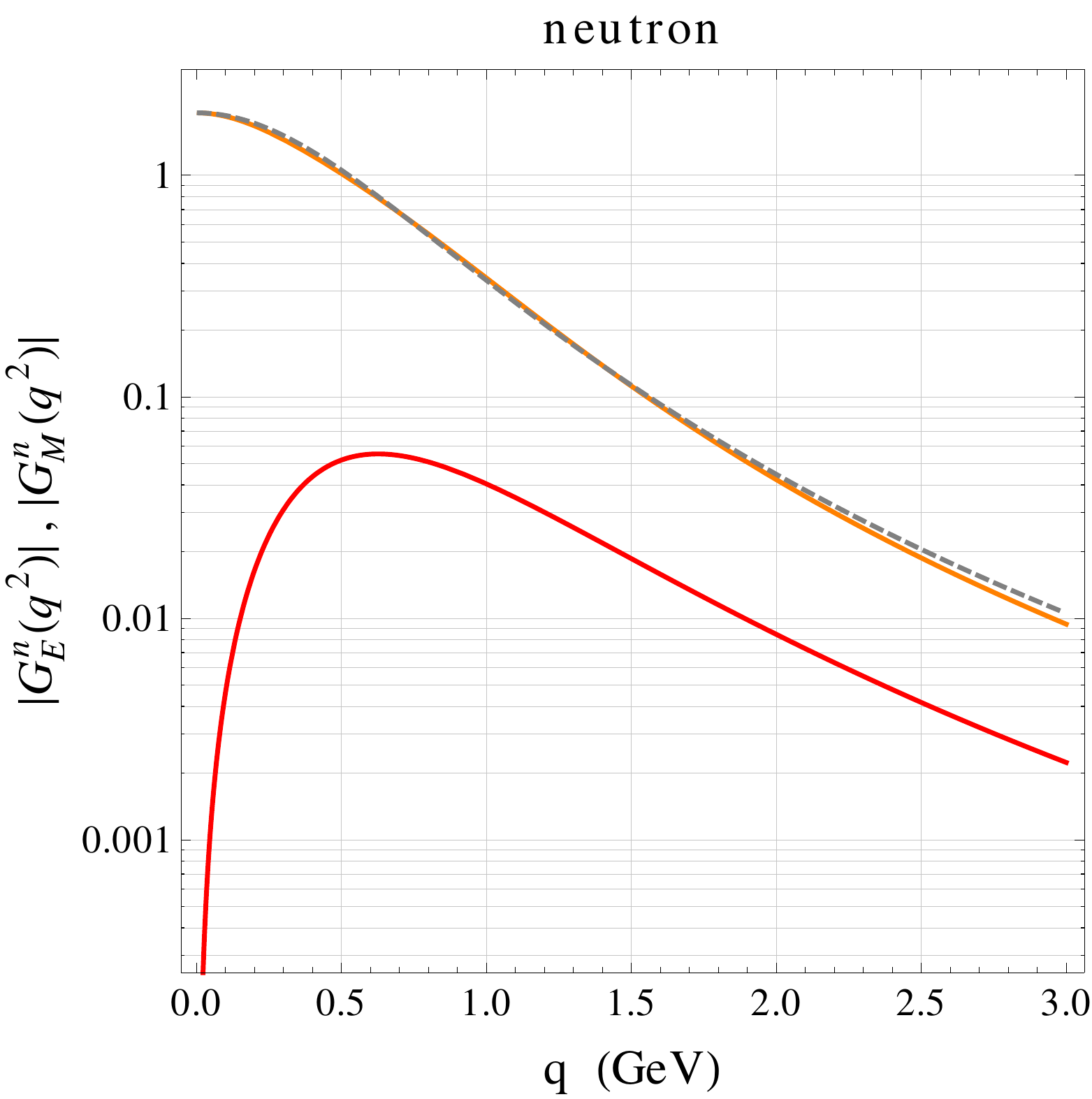} 
\caption{Electric (red) and magnetic (orange) form factors of the proton and neutron from~\cite{Alberico:2008sz}. For comparison, simple dipole form factors are shown with the dashed lines (in this case, the electric form factor of the neutron vanishes)}.
\label{fig:ff2}
\end{figure}

Fig.~\ref{fig:ff2} shows the electric (red) and magnetic (orange) form factors of the proton and neutron (\ref{eq:GEp})~-~(\ref{eq:GEn}). For comparison, the standard dipole form factors used in~\cite{Lovseth:1971vv,Ballett:2018uuc} are shown with the dashed lines. We see that the different sets of form factors do not differ appreciably at low momentum transfer. To estimate form factor uncertainties we computed the incoherent trident cross sections also with the standard dipole form factors and found few \% differences with respect to the calculation using the form factors in (\ref{eq:GEp})~-~(\ref{eq:GEn}). In view of the other uncertainties from nuclear modeling discussed in Sec.~\ref{sec:SM_Xsection}, this difference is insignificant.

\section{Borexino bound on the $L_\mu - L_\tau$ gauge boson} \label{app:borexino}

In this appendix we detail our treatment of the Borexino constraint shown in Fig.~\ref{fig:LmuLtau}.
The Borexino experiment measures the rate of low energy solar neutrinos that scatter elastically on electrons~\cite{Bellini:2011rx,Agostini:2017ixy}. The most precise measurement is obtained for $^7$Be neutrinos which have an energy of $E_\nu = 862$~keV. The good agreement of the measured scattering rate with the expectations from the Standard Model allows one to put bounds on non-standard contributions to the neutrino-electron scattering cross section. 

The neutrino scattering rate at Borexino is proportional to the neutrino-electron scattering cross section of the three neutrino flavors, weighted by their respective fluxes at the earth. For $^7Be$ solar neutrinos, the flux ratios at the earth are approximately
\begin{equation}
\phi_{\nu_e} :  \phi_{\nu_\mu} :  \phi_{\nu_\tau} \simeq 54\% : 20\% : 26\% ~,
\end{equation}
where we used the expressions from~\cite{Nunokawa:2006ms} and the latest neutrino mixing parameters from~\cite{Esteban:2016qun}.
The differential scattering cross sections can be written as~\cite{Marciano:2003eq}
\begin{eqnarray} \label{eq:dsigma_borexino}
\frac{d}{dy} \sigma(\nu_i e \to \nu_i e)_\text{SM} &=& \frac{G_F^2 m_e E_\nu}{8 \pi} \left[ \left(g^V_{iiee} - g^A_{iiee}\right)^2 + \left(g^V_{iiee} +g^A_{iiee}\right)^2 (1-y)^2 \right. \nonumber \\
&& \qquad\qquad\qquad \left. - \left( (g^V_{iiee})^2 - (g^A_{iiee})^2 \right) \frac{m_e}{E_\nu} y \right] ~,
\end{eqnarray}
where $y$ is the electron recoil energy $E_R$  normalized to the energy of the incoming neutrino $E_\nu$, $y = \frac{E_R}{E_\nu}$, and the relevant couplings are given in the SM by (cf. Tab.~\ref{tab:couplings}) $g^V_{eeee} = 1 + 4\sin^2\theta_W $, $g^A_{eeee} = -1  $, $g^V_{\mu\mu ee} = g^V_{\tau\tau ee} = -1 + 4\sin^2\theta_W$, and $g^A_{\mu\mu ee} = g^A_{\tau\tau ee} = 1$. Integrating over $y$ between the minimal recoil energy considered at Borexino $y_\text{min} \simeq 0.22$~\cite{Agostini:2017ixy} and the maximal value allowed by kinematics $y_\text{max} = \frac{2 E_\nu}{2 E_\nu + m_e} \simeq 0.77$ we find
\begin{equation}
\sigma(\nu_e e \to \nu_e e)_\text{SM} : \sigma(\nu_\mu e \to \nu_\mu e)_\text{SM} : \sigma(\nu_\tau e \to \nu_\tau e)_\text{SM} \simeq 4.7 : 1 : 1 ~.
\end{equation}
The $L_\mu - L_\tau$ gauge boson can contribute to the $\nu_\mu e \to \nu_\mu e$ and $\nu_\tau e \to \nu_\tau e$ scattering processes at the 1-loop level, through kinetic mixing between the $Z'$ and the SM photon. The kinetic mixing becomes relevant if the momentum transfer is small compared to the muon and tau masses, as is the case in the low energy neutrino scattering at Borexino.
We find that the $Z'$ contributions can be easily incorporated by making the following replacements in Eq.~(\ref{eq:dsigma_borexino})
\begin{eqnarray}
 g^V_{\mu\mu ee} \to g^V_{\mu\mu ee} - \frac{e^2 (g')^2}{6\pi^2} \log\left(\frac{m_\tau^2}{m_\mu^2}\right) \frac{v^2}{m_{Z'}^2 + 2m_e y E_\nu} ~, \\
 g^V_{\tau\tau ee} \to g^V_{\tau\tau ee} + \frac{e^2 (g')^2}{6\pi^2} \log\left(\frac{m_\tau^2}{m_\mu^2}\right) \frac{v^2}{m_{Z'}^2 + 2m_e y E_\nu}  ~. 
\end{eqnarray}
Contributions to the $\nu_e e \to \nu_e e$ process arise first at 2-loop and require $Z'$ mixing with the SM $Z$ boson. They are therefore negligible.
Note that the $Z'$ contributions to the muon-neutrino and tau-neutrino scattering differ by a relative minus sign. The new physics interferes constructively in $\nu_\mu e \to \nu_\mu e$ and destructively in $\nu_\tau e \to \nu_\tau e$, resulting in a partial cancellation of the new physics effect.

The change in the neutrino scattering rate at Borexino due to the presence of the $Z'$ can be determined as
\begin{equation}
\frac{\sigma_\text{Borexino}}{\sigma_\text{Borexino}^\text{SM}} = \frac{\sigma(\nu_e e \to \nu_e e)\phi_{\nu_e}+\sigma(\nu_\mu e \to \nu_\mu e)\phi_{\nu_\mu}+\sigma(\nu_\tau e \to \nu_\tau e)\phi_{\nu_\tau}}{\sigma(\nu_e e \to \nu_e e)_\text{SM} \phi_{\nu_e}+\sigma(\nu_\mu e \to \nu_\mu e)_\text{SM} \phi_{\nu_\mu}+\sigma(\nu_\tau e \to \nu_\tau e)_\text{SM} \phi_{\nu_\tau}}  ~. 
\end{equation}
Standard Model predictions for the scattering rate depend on the solar model, in particular on the assumed metallicity of the sun. Combining the predictions from~\cite{Vinyoles:2016djt} with the Borexino measurement in~\cite{Agostini:2017ixy}, we find the following range of allowed values for the scattering rate at the $2\sigma$ level: $0.88 < \sigma_\text{Borexino} / \sigma_\text{Borexino}^\text{SM} < 1.24$. This leads to the bound on the $Z'$ parameter space shown in Fig.~\ref{fig:LmuLtau}.

\end{appendix}

\bibliography{references}
\bibliographystyle{apsrev4-1}

\end{document}